\documentclass[preprint,review,12pt]{elsarticle}

\usepackage{amssymb}

\usepackage{hyperref}
\usepackage{amsmath}
\usepackage{cleveref}
\usepackage{booktabs}
\usepackage{caption,setspace}
\usepackage{tabularx}
\usepackage[export]{adjustbox}
\usepackage{tikz}
\usepackage{geometry}
\usepackage[T1]{fontenc}
\usepackage{wrapfig} 
\usepackage{flushend} 
\usepackage{helvet}  

\hypersetup{pdfauthor={Name}}

\makeatletter
\newcommand{\ssymbol}[1]{^{\@fnsymbol{#1}}}
\makeatother

\usepackage{array}
\newcolumntype{P}[1]{>{\centering\arraybackslash}p{#1}}

\AtBeginDocument{
    \crefformat{equation}{#2Eq.~#1#3}
    \crefformat{figure}{#2Fig.~#1#3}
    \crefformat{table}{#2Tab.~#1#3}
}

\journal{Computers in Biology and Medicine}  

\begin{document}
\begin{sloppypar}

\begin{frontmatter}

\title{Recent Progress in Transformer-based Medical Image Analysis}

\author[a]{Zhaoshan Liu}
\ead{e0575844@u.nus.edu} 

\author[a,b]{Qiujie Lv}
\ead{lvqj5@mail2.sysu.edu.cn} 

\author[a,b]{Ziduo Yang}
\ead{yangzd@mail2.sysu.edu.cn} 

\author[a]{Yifan Li}
\ead{e0576095@u.nus.edu} 

\author[c]{Chau Hung Lee}
\ead{chau_hung_lee@ttsh.com.sg}

\author[a]{Lei Shen\corref{cor1}}
\ead{mpeshel@nus.edu.sg} 
\cortext[cor1]{Corresponding author}

\address[a]{Department of Mechanical Engineering, National University of Singapore, 9 Engineering Drive 1, Singapore, 117575, Singapore}

\address[b]{School of Intelligent Systems Engineering, Sun Yat-sen University, No.66, Gongchang Road, Guangming District, 518107, China}

\address[c]{Department of Radiology, Tan Tock Seng Hospital, 11 Jalan Tan Tock Seng, Singapore, 308433, Singapore}

\begin{abstract}
The transformer is primarily used in the field of natural language processing. Recently, it has been adopted and shows promise in the computer vision (CV) field. Medical image analysis (MIA), as a critical branch of CV, also greatly benefits from this state-of-the-art technique. In this review, we first recap the core component of the transformer, the attention mechanism, and the detailed structures of the transformer. After that, we depict the recent progress of the transformer in the field of MIA. We organize the applications in a sequence of different tasks, including classification, segmentation, captioning, registration, detection, enhancement, localization, and synthesis. The mainstream classification and segmentation tasks are further divided into eleven medical image modalities. A large number of experiments studied in this review illustrate that the transformer-based method outperforms existing methods through comparisons with multiple evaluation metrics. Finally, we discuss the open challenges and future opportunities in this field. This task-modality review with the latest contents, detailed information, and comprehensive comparison may greatly benefit the broad MIA community.
\end{abstract}



\end{frontmatter}



\section{Introduction}
\label{1}

Transformer \citep{vaswani2017attention} is one of the most widely used models in the natural language processing (NLP) field and has achieved great success in many tasks, such as paraphrase generation \citep{egonmwan2019transformer}, text-to-speech synthesis \citep{chen2022fine}, and speech recognition \citep{shi2021emformer}. It is designed for transduction and sequence modeling with the remarkable capability of modeling long-range dependencies with the data. The convolutional-free transformer is based on the self-attention (attention) mechanism, a successful NLP technique \citep{lin2017structured, parikh2016decomposable, paulus2017deep, cheng2016long, abdulazeem2021cnn} that relates different positions of a single sequence to compute the sequence's representation \citep{vaswani2017attention}. Unlike the NLP field, the computer vision (CV) field has been dominated by the convolutional neural network (CNN) \citep{lecun1998gradient} for a long time \citep{kolesnikov2020big, xu2019spatiotemporal, lei2021breast, xie2021oriented, sun2021sparse}. Even if, many trials have been carried out to combine CNN and attention in the CV field \citep{wang2018non, carion2020end} while none of them overperform CNN. Until 2020, Dosovitskiy et al. \citep{dosovitskiy2020image} proposed a pioneering model and prove that implementing the transformer directly to sequences of image patches works well for image classification. In detail, the proposed method split the input image into multiple patches and embeds each of them linearly. With additional position embeddings added, the resulting vector sequences are fed to the transformer encoder. With the solid foundation they set, the transformer-based method has been widely adopted in the field of CV with superior performance \citep{chen2021crossvit, strudel2021segmenter, misra2021end, liu2023cect}.

Medical image analysis (MIA) is an essential branch in the CV field. Medical imaging utilizes various modalities to create a visual representation of the inside body \citep{saha2016active} and is of great help for further medical diagnosing. There are several kinds of medical imaging modalities, such as magnetic resonance imaging (MRI), computed tomography (CT), ultrasound (US), positron emission tomography (PET), optical coherence tomography (OCT), and digital fundus imaging (DFI). In practice, MIA is usually performed qualitatively by medical personnel. This may result in varying interpretations and degrees of accuracy because of varying degrees of reader experience or varying image quality. Moreover, such image analysis may be time- and labor-expensive. Due to these, the deep learning (DL) method has been widely applied in the field of MIA to reduce inter-reader variation as well as reduce time and manpower costs \cite{yang2021lung, poonkodi20233d, chen2022uncertainty, li2023transforming}. With the rapid development of the transformer in CV, the transformer-based method has been widely used in MIA either using the transformer solely \citep{le2022covid, krishnan2021vision, wu2021vision} or hybridizing CNN and transformer to capture both local and global information \citep{gu2022chest, duong2021detection, jiang2022multisemantic}. To help researchers catch up with this emerging research field, it is timely and important to have a comprehensive review and perspectives on transformer-based MIA.

\begin{figure*}
	\centering
		\includegraphics[width=0.88\textwidth]{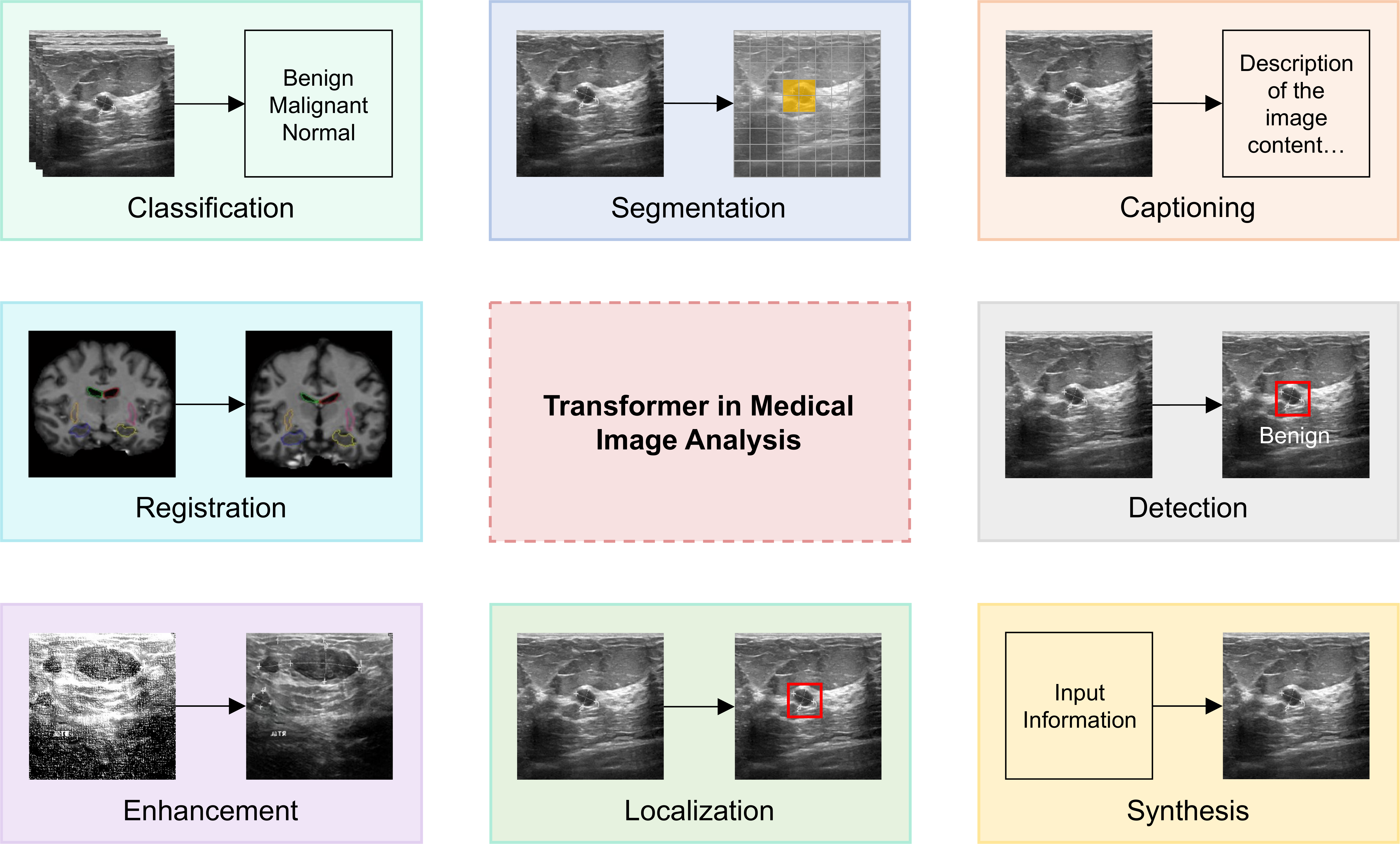}
	  \caption{MIA tasks included in this review. The tasks are organized in a sequence of classification, segmentation, captioning, registration, detection, enhancement, localization, and synthesis \citep{song2022td, al2020dataset}.}
	  \label{fig1}
\end{figure*}

\begin{figure*}
	\centering
		\includegraphics[width=0.96\textwidth]{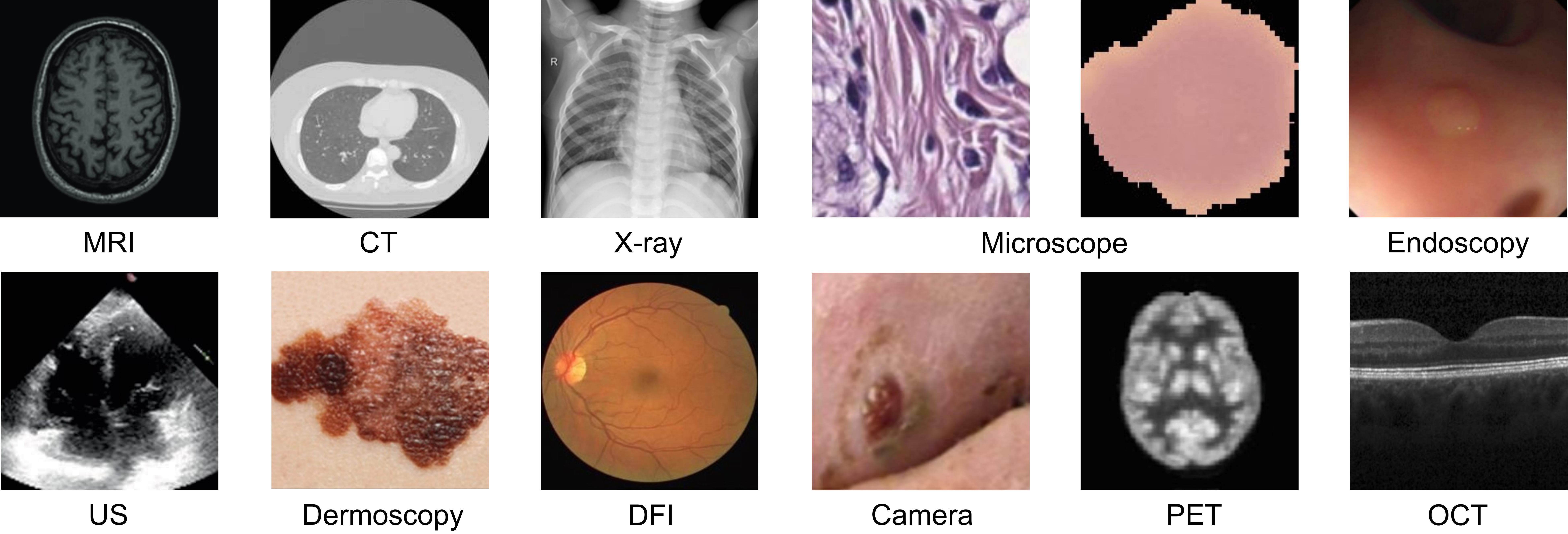}
	  \caption{Examples of modalities included in this work. Sequences are MRI \citep{hu2021cross}, CT \citep{morozov2020mosmeddata}, X-ray \citep{krishnan2021vision}, microscope \citep{kather2016multi} (left), \citep{islam2022explainable} (right), endoscopy \citep{silva2014toward}, US \citep{ouyang2020video}, dermoscopy \citep{aladhadh2022effective}, DFI \citep{chen2022pcat}, camera \citep{qayyum2021efficient}, PET \citep{zeng20223d}, and OCT \citep{kermany2018identifying}.}
	  \label{fig2}
\end{figure*}

In this review, we systematically introduce the transformer technique and its recent progress in the field of MIA, followed by outlooks and perspectives. We first recap the core component of the transformer, the attention mechanism, and the transformer itself. Then, we summarize the transformer-based applications in the sequence of different MIA tasks, including classification, segmentation, captioning, registration, detection, enhancement, localization, and synthesis, as shown in \Cref{fig1}. For the mainstream classification and segmentation tasks, we further divided their corresponding works into different medical imaging modalities. There are a total of eleven modalities in our review, including MRI, CT, X-ray, microscope, endoscopy, US, dermoscopy, DFI, camera, PET, and OCT, as shown in \Cref{fig2}. Finally, the open challenges and future research opportunities of transformer-based MIA tasks are also discussed. 

As this review is being written, we become aware of two similar review works \citep{he2022transformers, parvaiz2022vision}. The first review overviews the attention mechanism and other components essential for building the transformer. Then, the transformer architectures designed for MIA applications are discussed. The transformer-based applications are discussed task-wise, in which classification and segmentation applications are further divided into pure and hybrid transformer-based. Extensive discussions are then conducted, including learning manners, model improvements, and performance comparison with CNN. The second review starts by illustrating various medical imaging modalities, followed by an introduction to DL concepts, techniques, and architectures. The different tasks and the transformer-based applications are further discussed. Finally, research trends, current challenges, and future prospects are discussed. Here, we would like to highlight a couple of main differences between our review and others published. First, we cover more than 100 of the latest relevant papers, providing readers with a view of the latest research progress. We also highlight the latest transformer models leveraged in MIA, such as the swin transformer \citep{liu2021swin}, O-Net \citep{wang2022net}, and transformer-based region-edge aggregation network \citep{chen2023transformer}. The swin transformer is introduced in great detail as it is widely accepted and well-performed across different tasks and modalities. Next, our review contains more details. We include works with more than one modality and summarize them in a one-to-multi manner. We also summarize the objects researched, the datasets used, and the disease corresponding to these datasets when applicable. The summary of contents is coherently performed throughout different tasks. Sufficient details can help new researchers in the field grasp the necessary concepts easier and faster. Finally, instead of giving a quantitative performance evaluation of the transformer-based method, we also provide a comprehensive performance comparison between the transformer-based models and existing state-of-the-art DL methods in MIA. This proves the effectiveness of the transformer-based method. In summary, our task-modality review presents updated contents, detailed information, and comprehensive comparison that will greatly benefit the MIA community.

The rest of this review is organized as follows: In Section \hyperref[2]{2}, we show the methodology performed for our systematic review. In Section \hyperref[3]{3}, we recap the principle of the attention mechanism, the detailed structures of the transformer, and depict how the transformer is adopted into the MIA field. An introduction to different training manners and MIA tasks is also included. Section \hyperref[4]{4} organizes the transformer-based MIA applications from the perspective of different tasks. To better organize the large number of works related to mainstream classification and segmentation tasks, we further categorize them based on the imaging modalities. The objects in the relevant references are tabulated in detail. The datasets used as well as the diseases corresponding to the datasets are also tabulated when applicable. Moreover, a quantitative performance comparison across the transformer-based method and existing methods are summarized separately. In Section \hyperref[5]{5}, we point out the current challenges and future opportunities of the transformer-based MIA. A concise and comprehensive conclusion can be found in Section \hyperref[6]{6}.


\section{Methodology}
\label{2}

With the fast development of the transformer in the CV field, the research on the transformer has become one of the most popular research directions in the MIA field. We search on the Scopus database using "transformer" on the "title" field and "vision" and "medical" on the "title-abs-key" field and the results show that the number of papers in 2022 is more than four times compared to that of in 2021. What is more, the number of publications in 2023 is already more than that of 2021. We thus want to explore several research questions for the transformer-based MIA. First, on which MIA tasks have the transformer-based been successfully applied? Second, does the transformer outperforms previous DL methods, such as the CNN-based method? Finally, what are the current challenges or problems to be solved and the corresponding potential feasible solutions?

\begin{figure*}
	\centering
		\includegraphics[width=0.5\textwidth]{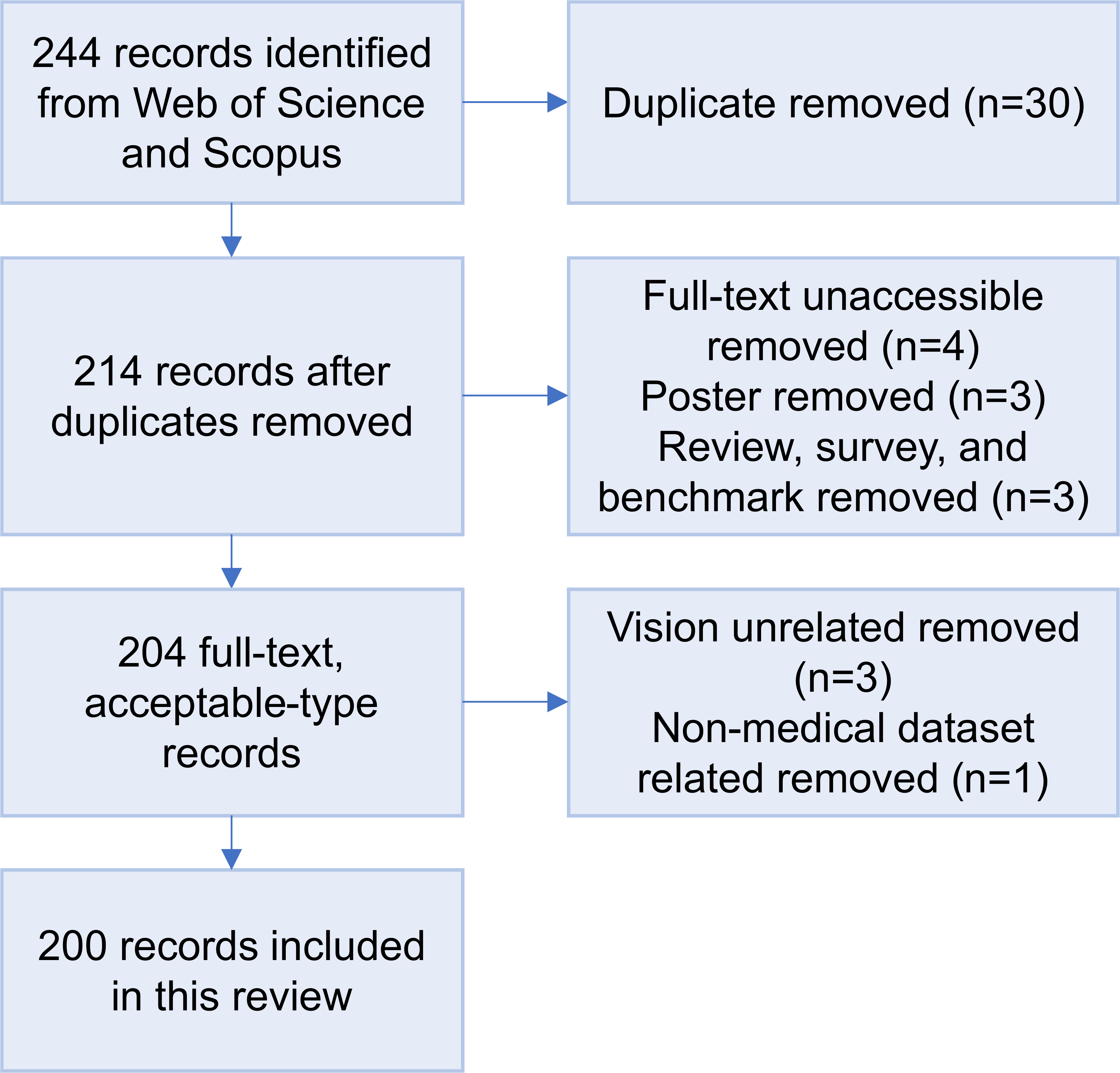}
	  \caption{Flow diagram for our systematic review.}
	  \label{fig3}
\end{figure*}

In this systematic review, the relevant references are collected by searching within the Web of Science and Scopus databases. The search timeframe is from 2021 to 8 Feb 2023. In the Web of Science database, we include keywords including "transformer", "classification", "segmentation", "captioning", "registration", "detection", "enhancement", "reconstruction", "denoising", "localization", "synthesis", "generation", and "diagnosis" in the "title" field. We also include the keywords "medical" and "vision" in the "topic" field. Regarding the Scopus database, we include keywords in the "TITLE" and "TITLE-ABS-KEY" fields identical to that of the "title" and "topic" fields, respectively. With papers searched, we then exclude the results that are duplicated. Then, we remove the records, in which full text is inaccessible. Conference posters, as well as review, survey, and benchmark papers, are also excluded. Note that the conference papers are not excluded. The papers submitted on arXiv are also reserved. Finally, we check the content of the papers to remove papers that are not vision related. We find that several papers are included even if we include the "vision" keyword. We also exclude the papers using non-medical datasets. Following PRISMA \cite{prisma}, we show the flow diagram for our systematic review in \Cref{fig3}.

\section{Background}
\label{3}

\subsection{Attention Mechanism}

The attention mechanism is the core component of the transformer. It differentially weights the significance of each part of the input data and allows the inputs to interact with each other to find to whom they should pay more attention. The attention mechanism expresses the importance of each input (e.g., token) in the current context as the attention score. The outputs are the aggregation of these interactions weighted by the corresponding attention scores. Specifically, with three attention vectors named query, keys, and values, the mechanism maps a query and a set of key-value pairs to output. The output is computed by summarizing all values according to the weight, which is calculated using a compatibility function of the query with the related key \citep{vaswani2017attention}. The attention mechanism can be divided into scaled dot-product attention and multi-head attention.

\textbf{Scaled dot-product attention}. Several queries, keys, and values compose the inputs of the scaled dot-product attention. The queries and keys have a dimensionality of $d_{k}$ and the values have a dimensionality of $d_{v}$. The dot product of all keys and the query is calculated. The resulting values are divided by a scale factor, $\sqrt{d_{k}}$, and pass a softmax function. The attention can be calculated through the dot product with the values. Practically, a set of queries, keys, and values are packed into corresponding matrices, \emph{Q}, \emph{K}, \emph{V}, and the outputs matrix can be calculated using the below formula:
$$
\operatorname{Attention}(Q, K, V)=\operatorname{softmax}\left(\frac{Q K^{T}}{\sqrt{d_{k}}}\right) V
$$

\begin{figure*}
	\centering
		\includegraphics[width=0.66\textwidth]{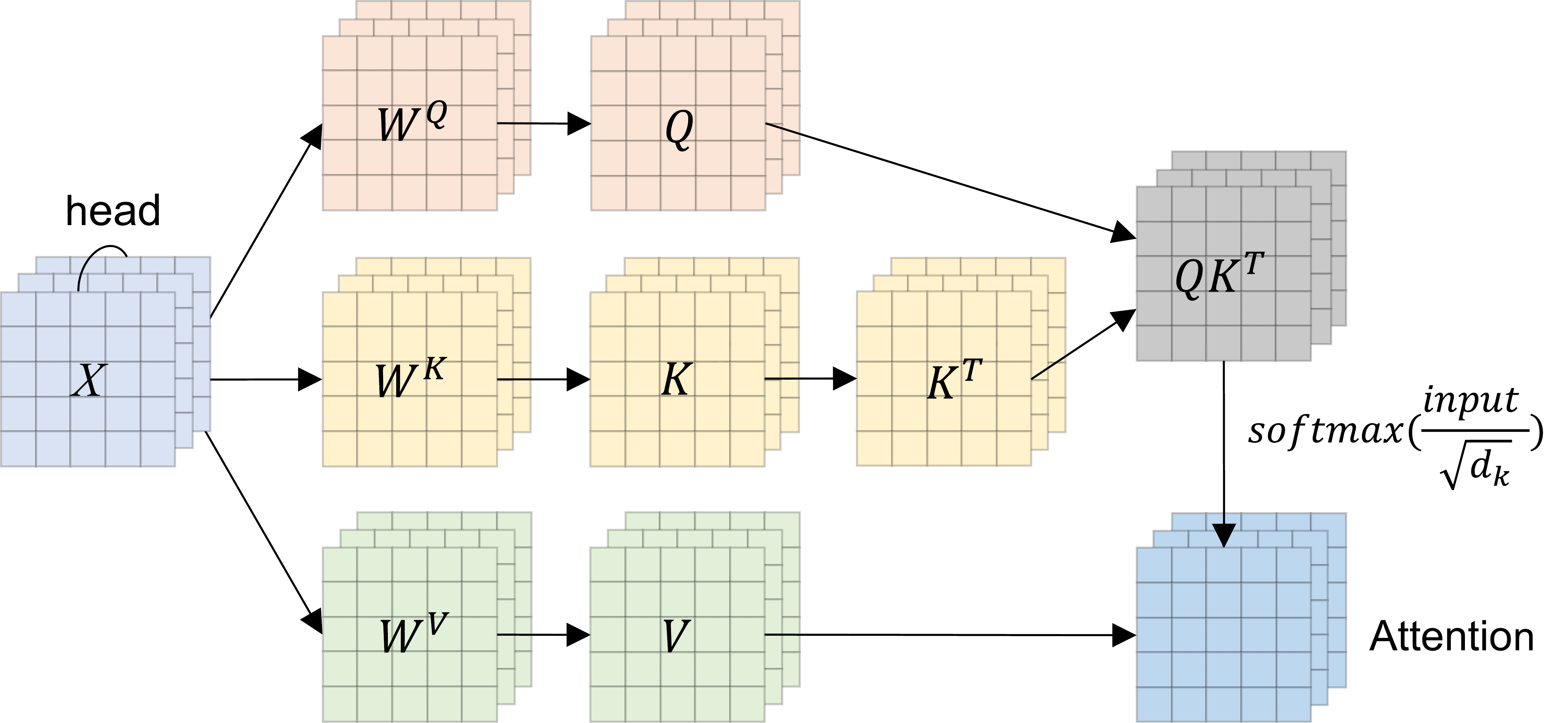}
	  \caption{An intuitive illustration of the multi-head attention mechanism. The result of each head is calculated respectively, and the values are then concatenated. \emph{X} denotes the input, $W^Q$, $W^K$, and $W^V$ represent the parameter matrices for projections, and $\sqrt{d_{k}}$ stands the scale factor.}
	  \label{fig4}
\end{figure*}

\textbf{Multi-head attention}. The scaled dot-product attention is single-head attention. In practice, multi-head attention is more often used as it improves the expressive power of the model and stabilizes training. The multi-head attention allows the model to jointly attend to information from different representation subspaces at different positions. Specifically, the queries, keys, and values are projected for \emph{h} times, where \emph{h} is the number of heads. The attention function is performed on each of the projected results concurrently. The output values are concatenated and projected again to obtain the final results \citep{vaswani2017attention}. A concise illustration of the multi-head attention is shown in \Cref{fig4}, and the multi-head attention can be described using the below equation:
$$
\begin{aligned}
\operatorname{MultiHead}(Q, K, V) &=\operatorname{Concat}\left(\operatorname{head}_{1}, \ldots, \operatorname{head}_{\mathrm{h}}\right) W^{O} \\
\text {where head}_{\mathrm{i}} &=\operatorname{Attention}\left(Q W_{i}^{Q}, K W_{i}^{K}, V W_{i}^{V}\right)
\end{aligned}
$$
where $W_{i}^{Q} \in \mathbb{R}^{d_{\text {model }} \times d_{k}}, W_{i}^{K} \in \mathbb{R}^{d_{\text {model }} \times d_{k}}, W_{i}^{V} \in \mathbb{R}^{d_{\text {model }} \times d_{v}}$, $W^{O} \in \mathbb{R}^{h d_{v} \times d_{\text {model }}}$ are the parameter matrices for projections, and $d_{\text {model }}$ is dimensional keys, values, and queries.

\begin{figure*}
	\centering
		\includegraphics[width=0.54\textwidth]{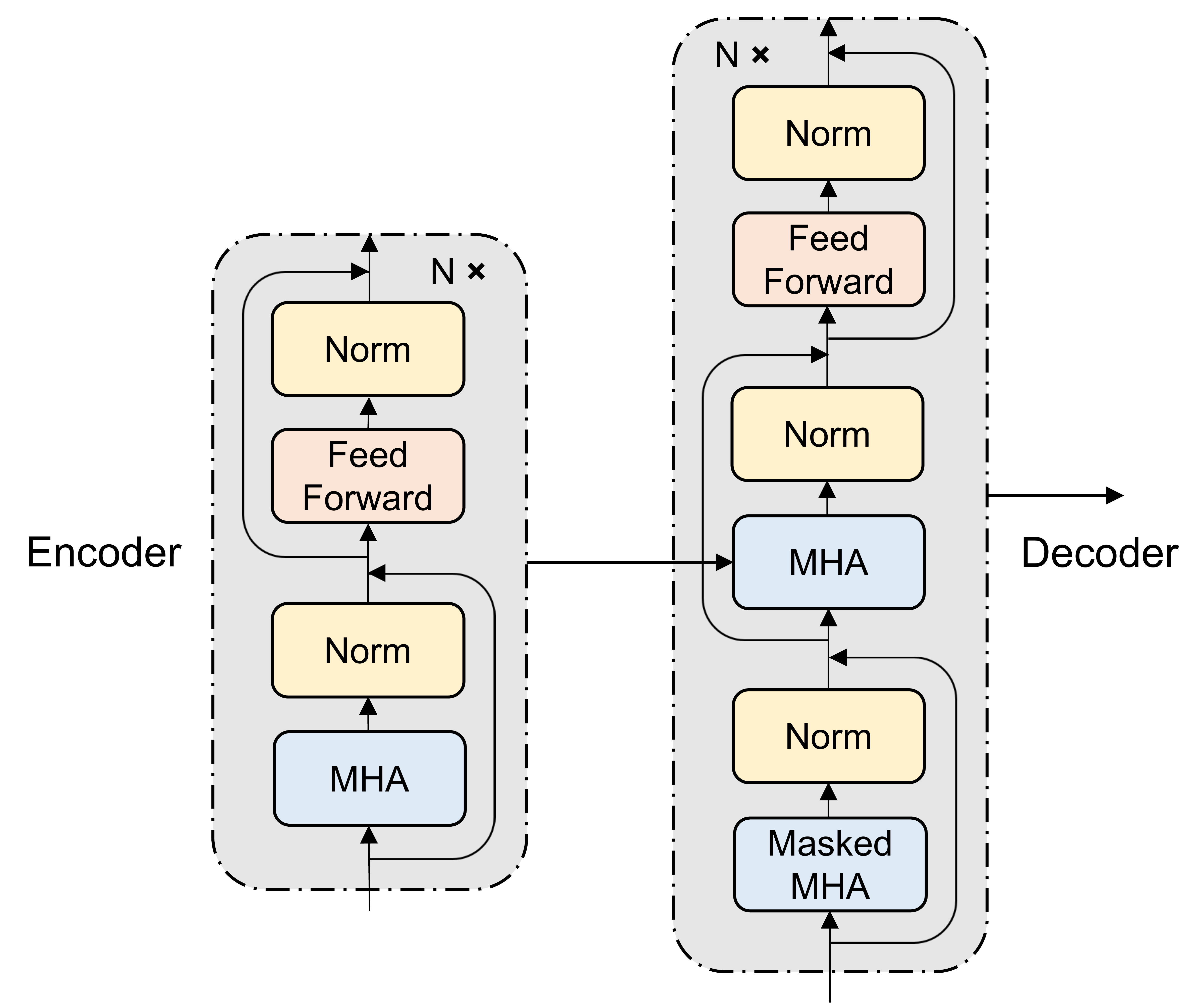}
	  \caption{The encoder-decoder structure in transformer \citep{vaswani2017attention}. MHA refers to the multi-head attention module and Norm represents layer normalization. The output of the encoder is fed into the MHA in the decoder. \textbf{Left}: encoder, and \textbf{Right}: decoder.}
	  \label{fig5}
\end{figure*}

\subsection{Transformer}

Researchers have proposed several variant models based on the attention mechanism, which generally combines the attention mechanism with the recurrent neural network (RNN), such as LSTM. These models are usually limited in training speed due to the sequential structure, and the parallel computing ability is limited. Since the attention model itself can capture global information, a natural question raised is whether we can remove the RNN structure and rely only on the attention model. The answer is yes. The transformer is such a novel model utilized to address sequence-to-sequence tasks, taking a sequence as the input and generating the predicted probabilities as the output. It is mainly composed of the attention mechanism and has an encoder-decoder structure, as shown in \Cref{fig5}. Both the encoder and decoder are tandem by several identical blocks. The blocks are composed of several parts, including the masked multi-head attention module, multi-head attention module, layer normalization, and position-wise feed-forward network. The masked multi-head attention module employs the attention mechanism up to the current position and excluded the unpredicted positions till now. Combining the masked multi-head attention and the position offset of the output embeddings, the predictions for the position only lie on the outputs at earlier positions can be ensured \citep{vaswani2017attention}. The fully connected feed-forward network consists of two linear layers and a ReLU activation function in between. The procedure can be expressed using the below equation:
$$
\operatorname{FFN}(x)=\max \left(0, x W_{1}+b_{1}\right) W_{2}+b_{2}
$$

\textbf{Encoder}. The transformer encoder consists of \emph{N} identical blocks. For each block, there are two main parts and both parts employ the residual connection proposed by He et al. \citep{he2016deep}. The bottom part is composed of a multi-head attention module and layer normalization. The top part consists of a fully connected feed-forward network and layer normalization.

\textbf{Decoder}. The transformer decoder is composed of \emph{N} identical blocks like the encoder and consists of three main parts in each block. The bottom part is composed of a masked multi-head attention module and layer normalization. The middle part consists of a multi-head attention module followed by layer normalization. It is worth noting that the multi-head attention module in the decoder also takes the encoder's output as the input. The top part is composed of a feed-forward network and layer normalization. The residual connection is implemented for all three parts.

\begin{figure*}
	\centering
		\includegraphics[width=0.96\textwidth]{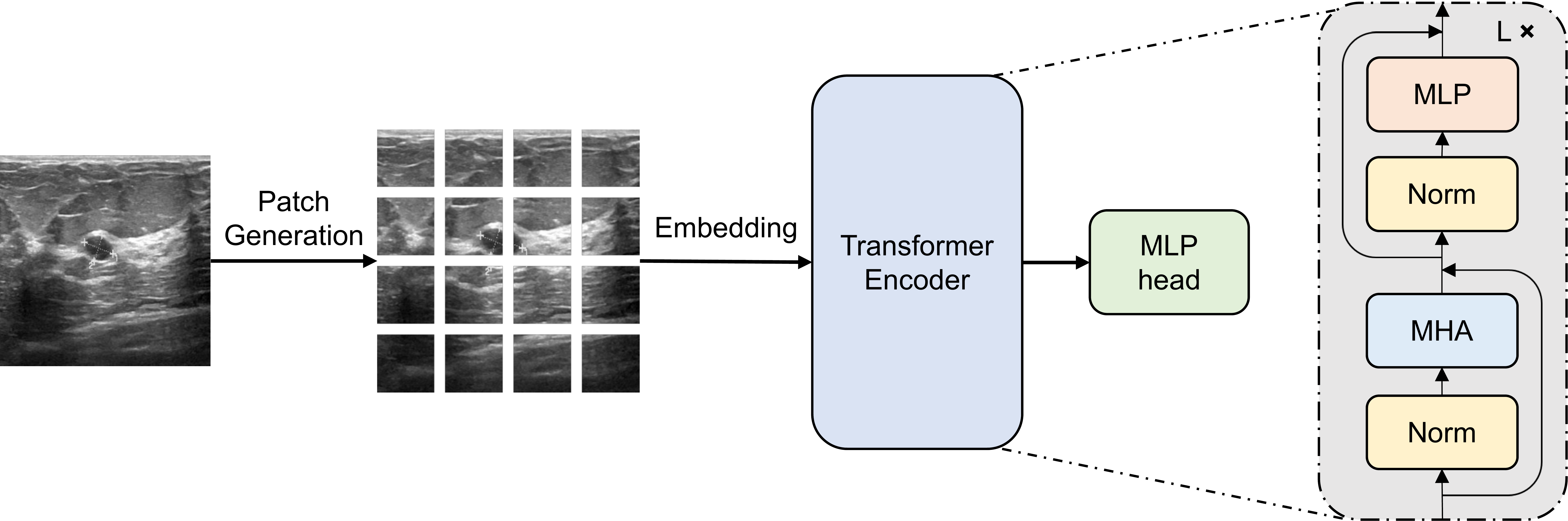}
	  \caption{Detailed structure of the transformer used in the MIA field. MHA refers to the multi-head attention module. Norm represents layer normalization and MLP illustrates the multilayer perception module.}
	  \label{fig6}
\end{figure*}

\subsection{Transformer in MIA}

To employ the transformer in the field of MIA, the input medical images need to be pre-processed \citep{dosovitskiy2020image} as they are 2D. The pre-processing process can be divided into two main steps, patch generation, and embedding. The patch generation splits the input image into several patches, and the embedding flattens the patches and generates patch embeddings. Positional embeddings and class embedding are also added. The processed images are fed into the transformer encoder for feature extraction. The transformer encoder undergoes slight modification by altering the sequence of layer normalization. The output from the transformer encoder serves as the input for the multilayer perceptron head, resulting in the image class prediction. The detailed structure of the transformer used in MIA is illustrated in \Cref{fig6}.

\textbf{Patch generation}. The patch generation receives image $\mathbf{x} \in \mathbb{R}^{H \times W \times C}$ as the input, where \emph{H} and \emph{W} are the height and the weight of the image, respectively. The input image is reshaped into a patch sequence $\mathbf{x}_{p} \in \mathbb{R}^{N \times\left(P^{2} \cdot C\right)}$, where \emph{P} is the dimension of each image patch, and \emph{N} is the number of patches \citep{dosovitskiy2020image}. The relationship between \emph{N}, \emph{H}, \emph{W}, and \emph{P} can be expressed as:
$$
N=H W / P^{2}
$$

\textbf{Embedding}. The embedding obtains patches as the input. It flattens and maps them to \emph{D} dimensions using a linear projection $\mathbf{E}$, where \emph{D} is the latent vector size of the transformer layers. The resulting patch embeddings are concatenated with the class embedding \citep{devlin2018bert} and then the concatenated embeddings are summed with the position embeddings to retain positional information. The embedding can be described through the below equation \citep{dosovitskiy2020image}:
$$
\mathbf{y}=\left[\mathbf{x}_{\text{class}} ; \mathbf{x}_{p}^{1} \mathbf{E} ; \mathbf{x}_{p}^{2} \mathbf{E} ; \cdots ; \mathbf{x}_{p}^{N} \mathbf{E}\right]+\mathbf{E}_{\text {pos }} \quad \text{where} \ \mathbf{E} \in \mathbb{R}^{\left(P^{2} \cdot C\right) \times D}, \mathbf{E}_{p o s} \in \mathbb{R}^{(N+1) \times D}
$$

\textbf{Transformer encoder}. The transformer encoder takes the resulting embeddings as the input. The overall encoder structure is similar to that in \Cref{fig5} while the layer normalization is moved before the multi-head attention module and multilayer perceptron module (feed-forward network). The multilayer perceptron module is composed of two linear layers and both of the layers use GELU as the activation function.

\textbf{Multilayer perceptron head}. The output of the transformer encoder is fed into the multilayer perceptron head to get the classification result. At the pre-training stage, the multilayer perceptron head is composed of one hidden layer, while at the fine-tuning stage, it consists of a single linear layer.

\begin{figure*}
	\centering
		\includegraphics[width=0.96\textwidth]{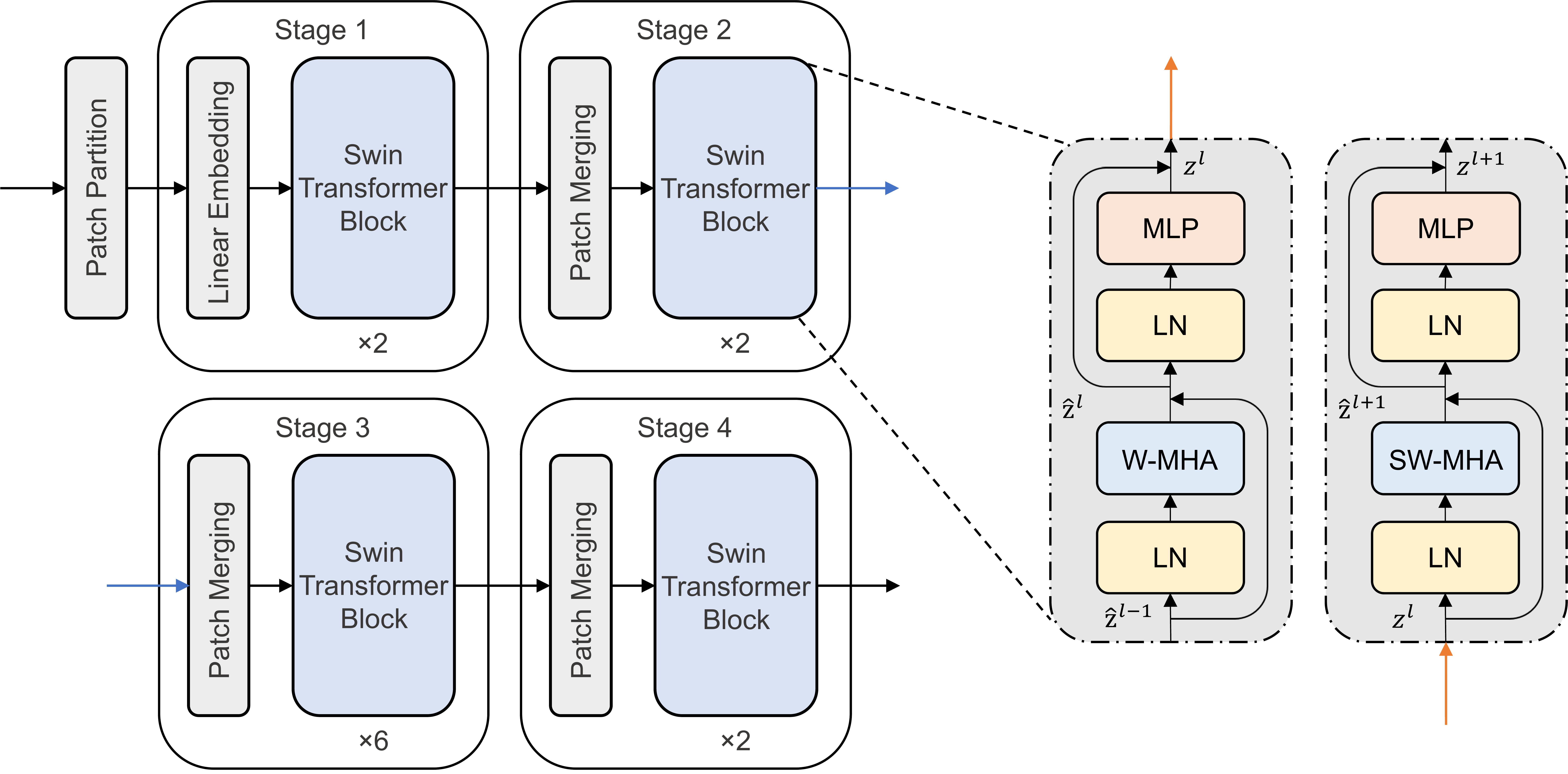}
	  \caption{The detailed structure of the swin transformer. The swin transformer consists of four stages and two successive swin transformer blocks form a basic unit. MHA, W-MHA, and SW-MHA denote the multi-head attention module, multi-head attention module with regular windowing configurations, and multi-head attention module with shifted windowing configurations, respectively. MLP represents the multilayer perception module, and LN is layer normalization. $\hat{\mathbf{z}}^l$ and $\mathbf{z}^l$ are the output features of the W-MHA and the MLP inside block $l$, respectively. Blue and orange arrows mean the connection of two parts.}
	  \label{fig7}
\end{figure*}

The transformer faces several challenges in the CV field that stem from the differences between the CV and NLP fields. For one thing, the scale of visual entities can vary largely. For another, the resolution of images can be very high. To solve these problems, the swin transformer has been proposed. The swin transformer builds hierarchical feature maps and calculates the attention only within each local window. Compared with computing attention globally, the computation complexity is reduced largely from quadratic to linear. To provide connections between different windows, a shifted window approach is proposed. The window partitioning is shifted after the attention is calculated within each window. With such designs, the swin transformer can not only reduce the computation complexity to linear from quadratic but also model at various scales with high flexibility. The overall structure of the swin transformer is shown in \Cref{fig7}. There are four stages in the swin transformer and two successive swin transformer blocks form a basic unit. The former swin transformer block has a multi-head attention module with regular windowing configurations, while the latter consists of a shifted windowing multi-head attention module. The multilayer perceptron module consists of two layers with a GELU activation function in between, and layer normalization is implemented in both of the blocks.

There are distinct advantages of using the transformer model over CNN in the field of MIA. In the MIA field, many involved images have repetitive patterns or symmetrical patterns, such as the microscope image (left) and the CT image shown in \Cref{fig2}. CNN is not sensitive to such symmetrical or repetitive patterns as it only focuses on local features. In contrast, the transformer can capture these global features by exploring the relationships among local regions. Take a tumor image consisting of repetitive normal patterns and a tumor pattern as an example, the calculated attention scores may show high similarities among repeatedly normal regions and low similarities between the tumor region and the normal regions. This can demonstrate that the transformer is sensitive to repetitive patterns.

The scarcity of datasets is a potential issue for the transformer. Training the transformer requires a large amount of data as it lacks some of CNN's inherent inductive biases, such as locality and translation equivariance. It is reported that training the transformer with less than 100 million images usually obtains a suboptimal solution compared to CNN, while the performance of the transformer continues increasing with the increase of the dataset size. However, the process of creating a large dataset of medical images is substantially different from nature images due to various reasons. For instance, it heavily relies on costly equipment to capture medical images, which subsequently requires human experts for annotation. Additionally, medical datasets cannot always be made publicly available due to patient privacy concerns. Therefore, collecting a dataset with more than 100 million images is a significant challenge in the MIA field, and sometimes there are only thousands or even hundreds of images in a dataset \citep{guo2021multi, tang2021recurrent}. In the case of common data shortage, data augmentation methods are widely implemented in MIA, such as traditional image transformations like flip \citep{al2019deep}, or newer image synthesis methods \citep{liu2022semi, zhao2019data, dorjsembe2022three}. Besides, transfer learning is another widely used technique in the MIA field \citep{chen2019med3d, zhou2017using, liu2022semi}. Transfer learning allows the model to be pre-trained on a larger dataset, such as ImageNet, and the learned knowledge from the bigger dataset can be utilized when training the smaller ones.

\subsection{Training Manner}

Similar to other DL models, the transformer is trained in different training manners. This can include supervised (full-supervised), unsupervised, semi-supervised, self-supervised, weakly supervised, et al. The supervised learning manner is the most widely implemented learning manner and data used for supervised learning contain full labels for model evaluation. Opposite to supervised learning, unsupervised learning only receives the input data and learns intrinsic data properties by discovering underlying structures, patterns, or relationships in the data. Weakly-supervised learning is a learning method between supervised learning and unsupervised learning, in which the learning algorithm is trained using incomplete, imprecise, or noisy labeled data. Semi-supervised learning is a combination of supervised learning and unsupervised, which uses a large amount of unlabeled data and a small amount of labeled data for training to improve model performance. Self-supervised learning enables the model to learn richer and more effective feature representations by designing specific pretext tasks and using the input data to generate supervisory signals. The two most widely implemented self-supervised learning algorithms are DINO \citep{caron2021emerging} and BYOL \citep{grill2020bootstrap}, where DINO utilizes the knowledge distillation technique and BYOL minimizes the difference in latent representations between augmented image pairs. The BYOL is composed of two networks, which are the online network and the target network. The target network is utilized to provide targets for online network training. The introduction of learning methods outside of supervised learning can reduce the requirement of the label amount to a large extent and has been widely implemented in the MIA field.

\subsection{MIA Task}

There are numerous tasks in the field of MIA. Here, we include classification, segmentation, captioning, registration, detection, enhancement, localization, and synthesis in our review. Classification is the process of categorizing images into distinct classes. Segmentation partitions images into various objects or subgroups and can be regarded as pixel-level classification. Captioning generates descriptive language using visual information. Registration involves transforming multiple sets of data obtained from different sensors, viewpoints, etc. \citep{brown1992survey} into a unified coordinate system. Object detection predicts the boundary and the classification result across different objects. It is worth noting that one type of object detection \citep{he2017mask} also performs pixel-level classification, while it is not widely implemented in the MIA field partly due to computation resource considerations. Localization is a similar task to object detection while it predicts the object boundary solely. Enhancement works to enhance patterns and remove noise artifacts, which primarily includes reconstruction and denoising. Reconstruction enhances image quality by addressing potential low signal-to-noise ratio, contrast-to-noise ratio, and artifacts \citep{singh2012medical}, while denoising enhances visual images through noise removal. Synthesis creates desired images, which is the opposite of classification. A concise graphical illustration for each task can be found in \Cref{fig1}.


\section{Applications}
\label{4}

\begin{figure*}
	\centering
		\includegraphics[width=0.96\textwidth]{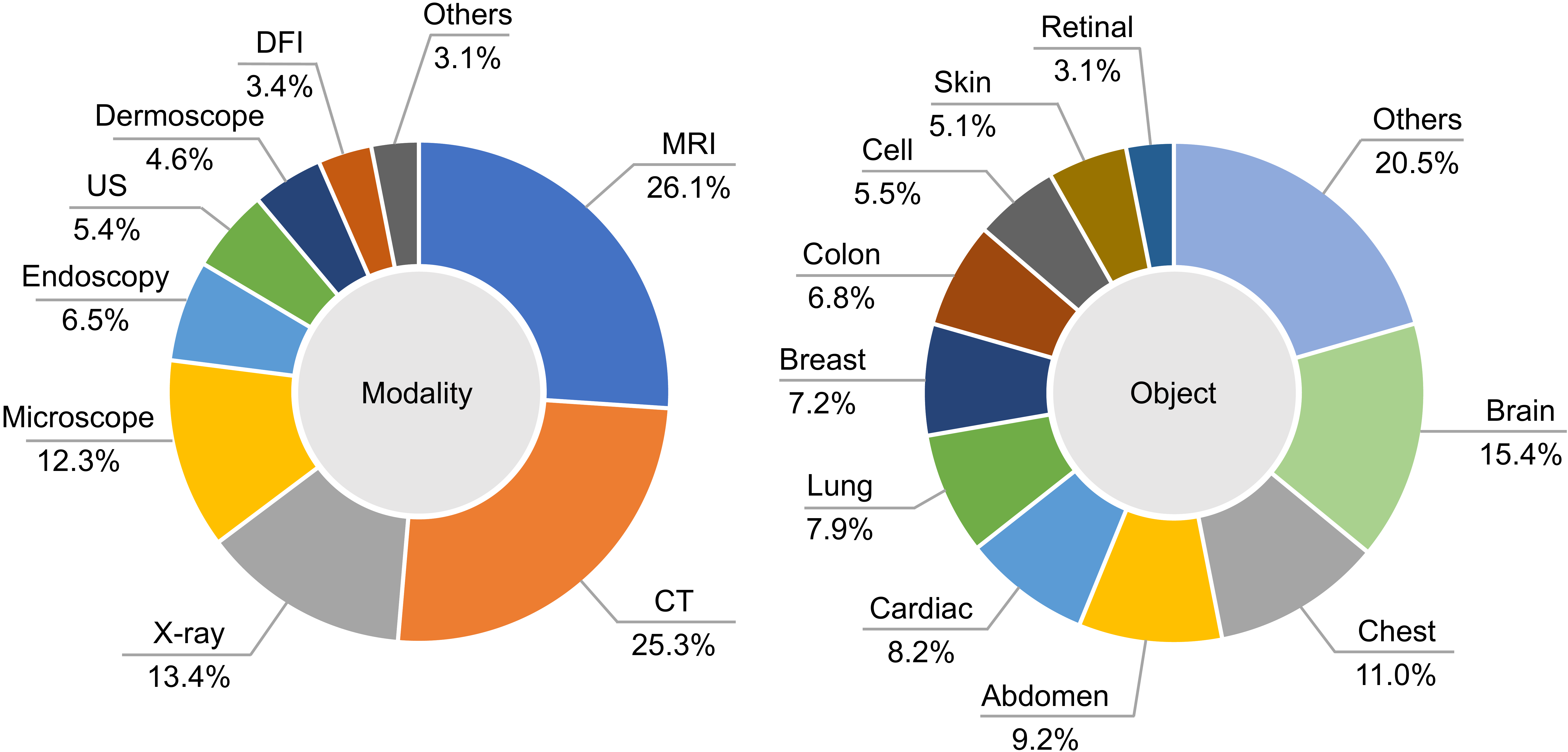}
	  \caption{The proportion that each modality and object. \textbf{Left}: modality, and \textbf{Right}: object. There are around 30 objects in "Others", and each one has a small proportion.}
	  \label{fig8}
\end{figure*}

We discuss the applications of the transformer in classification, segmentation, captioning, registration, detection, enhancement, localization, and synthesis tasks. Given the large number of papers involved in mainstream classification and segmentation tasks, we further categorize them by modalities. There are eleven modalities included in this work, which are MRI, CT, X-ray, microscope, endoscopy, US, dermoscopy, DFI, camera, PET, and OCT. It is worth noting that endoscopy includes colonoscopy, laryngoscopy, etc, and the whole slide image (WSI) is included in the microscope as it is also referred to as virtual microscope \citep{pantanowitz2011review}. Besides, magnetic resonance angiography is included in the MRI. The objects contained in the selected references are tabulated in detail. The proportion of each modality and object in all cited references are summarized in \Cref{fig8}. In order to facilitate statistical analysis, we present complex objects in a "grouped" manner, whereby several small objects are grouped into the same category. For instance, the colon covers the rectum. Additionally, the related datasets corresponding to the works are summarized in the tables, and the diseases with respect to the datasets are also tabulated when applicable. Diseases are also organized in a “grouped” way when appropriate. For example, lung disease can cover tuberculosis, COVID-19, pneumonia, pneumothorax, etc. The main narrative sequence within each section is that we usually start from the works using existing transformer models followed by the works with newly developed models. It is worth noting that we discuss the qualitative results like segmented masks within the applications for the sake of coherence, while the quantitative performance comparison is illustrated separately in Section \hyperref[4.4]{4.4}. This can make the performance comparison between transformer-based models and existing models more intuitive.

\subsection{Classification}
\label{4.1}

Since classification is one of the most widely-studied applications in MIA, we organize this sub-section in the sequence of different modalities, including X-ray, microscope, CT, MRI, DFI, dermoscopy, endoscopy, US, PET, camera, and OCT. In the case of papers containing more than one modality, we put them into the "multiple" category following the OCT. \Cref{tab1} summarizes the classification applications with the transformer. In the classification application, most of the works combine the transformer with CNN to capture both local and global information for further robustness and performance improvement.

\begin{table}
\centering
\caption{Transformer-based classification applications. The mark "-" shows the corresponding information is not publicly available.}
\resizebox{\linewidth}{!}{
\begin{tabular*}{708pt}{ m{7cm}<{\centering} c m{4cm}<{\centering} m{4.8cm}<{\centering} m{6.4cm}<{\centering} }
\toprule
    Method                                                                    & Year  & Modality                                             & Object                                                 & Dataset \\
\midrule
    transformer \citep{ukwuoma2022automated}                                  & 2022  & X-ray                                                & lung                                                   & lung disease \citep{badawi2021detecting, chowdhury2020can} \\
    swin transformer \citep{le2022covid}                                      & 2022  & X-ray                                                & chest                                                  & lung disease \citep{covidx, kermany2018labeled, tsai2021rsna, cohen2021radiographic} \\
    transformer \citep{krishnan2021vision}                                    & 2021  & X-ray                                                & chest                                                  & lung disease \citep{chowdhury2020can, rahman2021exploring} \\
    transformer, DeiT \citep{behrendt2022data}                                & 2022  & X-ray                                                & chest                                                  & \citep{irvin2019chexpert} \\
    transformer \citep{ayana2023vision}                                       & 2023  & X-ray                                                & breast                                                 & breast cancer \citep{bmim} \\      
    DeiT \citep{jalalifar2022data}                                            & 2022  & X-ray                                                & chest                                                  & - \\
    DeiT \citep{chen2022transformers}                                         & 2022  & X-ray                                                & breast                                                 & breast cancer \citep{zheng2012computer} \\  
    ViT-Eff \citep{duong2021detection}                                        & 2021  & X-ray                                                & chest                                                  & lung disease \citep{jaeger2014two, cohen2020covid, wang2017chestx, carion2020end} \\
    PneuNet \citep{wang2023pneunet}                                           & 2023  & X-ray                                                & lung                                                   & lung disease \citep{qatacovid, covidir, covideurorad, covidcxr, covidsirm, covidradio, covidcxnet, covidrsna, chestpneumonia, bustos2020padchest} \\
    Chest L-Transformer \citep{gu2022chest}                                   & 2022  & X-ray                                                & chest                                                  & lung disease \citep{filice2020crowdsourcing} \\ 
    MXT \citep{jiang2022mxt}                                                  & 2022  & X-ray                                                & chest                                                  & chest disease \citep{chestx}, \citep{catheter} \\
    multi-feature fusion transformer \citep{qi2022multi}                      & 2022  & X-ray                                                & chest                                                  & lung disease \citep{vaya2020bimcv, wang2020covid, desai2020chest, tsai2021rsna, clark2013cancer, covidir} \\
    MP-ViT \citep{jiang2022multisemantic}                                     & 2022  & X-ray                                                & chest                                                  & lung disease \citep{kermany2018identifying} \\
    federated split transformer \citep{park2021federated}                     & 2021  & X-ray                                                & lung                                                   & lung disease \citep{vaya2020bimcv, signoroni2021bs, borghesi2020covid} \\
    transformer \citep{ikromjanov2022whole}                                   & 2022  & Microscope                                           & prostate                                               & prostate cancer \citep{prostatewsi} \\
    transformer, compact convolutional transformer \citep{zeid2021multiclass} & 2021  & Microscope                                           & colon                                                  & colorectal cancer \citep{kather2016multi} \\
    explainable transformer-based \citep{islam2022explainable}                & 2022  & Microscope                                           & cell                                                   & malaria parasite \citep{rajaraman2018pre, fuhad2020deep} \\
    ensembled swin transformer \citep{tummala2022breast}                      & 2022  & Microscope                                           & breast                                                 & breast tumor \citep{spanhol2015dataset} \\
    IMGL-VTNet \citep{barmpoutis2022multi}                                    & 2022  & Microscope                                           & gastric                                                & gastric intestinal metaplasia \citep{imgl} \\
    AMIL-Trans \citep{zhang2022attention}                                     & 2022  & Microscope                                           & breast                                                 & breast cancer \citep{bejnordi2017diagnostic} \\
    Self-ViT-MIL \citep{gul2022histopathological}                             & 2022  & Microscope                                           & breast                                                 & breast cancer \citep{bejnordi2017diagnostic} \\
    TransPath \citep{wang2021transpath}                                       & 2021  & Microscope                                           & breast, colon                                          & breast cancer \citep{bejnordi2017diagnostic}, colorectal cancer \citep{kather2019predicting}, polyps \citep{wei2021petri} \\
    Fourier ViT \citep{duan2022fourier}                                       & 2022  & Microscope                                           & breast                                                 & breast cancer \citep{amgad2019structured} \\
    RAMST \citep{lv2022joint}                                                 & 2022  & Microscope                                           & gastrointestinal                                       & - \\
    CWC-Transformer \citep{wang2023cwc}                                       & 2023  & Microscope                                           & breast, lung                                           & breast cancer \citep{bejnordi2017diagnostic}, breast cancer\citep{srinivas2021bottleneck} \\
    transformer \citep{gai2022using}                                          & 2022  & CT                                                   & lung                                                   & lung disease \citep{rahimzadeh2021fully, morozov2020mosmeddata} \\
    transformer \citep{sufian2022pre}                                         & 2022  & CT                                                   & lung                                                   & lung disease \citep{soares2020sars} \\
    transformer \citep{li2021medical}                                         & 2021  & CT                                                   & lung                                                   & - \\
    transformer \citep{salvi2022vision}                                       & 2022  & CT                                                   & artery                                                 & - \\
    transformer \citep{sahoo2022vision}                                       & 2022  & CT                                                   & lung                                                   & lung disease \citep{soares2020sars} \\
    transformer \citep{wu2021vision}                                          & 2021  & CT                                                   & lung                                                   & lung disease \citep{emphysema, sorensen2010quantitative} \\
    multi-view convolutional transformer \citep{xiong2022pulmonary}           & 2022  & CT                                                   & lung                                                   & - \\
    DenseTransformer \citep{mei2022marrying}                                  & 2022  & CT                                                   & lung                                                   & lung disease \citep{zhao2020covid} \\
    transformer-based factorized encoder \citep{huang2022transformer}         & 2022  & CT                                                   & lung                                                   & lung disease \citep{afshar2021covid} \\
    multi-granularity dilated transformer  \citep{wu2022multi}                & 2023  & CT                                                   & lung                                                   & lung disease \citep{armato2011lung} \\
    transformer \citep{salanitri2022neural}                                   & 2022  & MRI                                                  & pancreas                                               & intraductal papillary mucosal neoplasms \citep{lalonde2019inn} \\
    TransMed \citep{dai2021transmed}                                          & 2021  & MRI                                                  & head, neck, knee                                       & anterior cruciate ligament, meniscal tears \citep{bien2018deep} \\
    double-scale GAN \citep{hu2021cross}                                      & 2021  & MRI                                                  & brain                                                  & \citep{iximri} \\
    MEST \citep{liu2022mest}                                                  & 2022  & MRI                                                  & brain                                                  & Parkinson’s disease \citep{marek2011parkinson} \\
    MIL-VT \citep{yu2021mil}                                                  & 2021  & DFI                                                  & retinal                                                & retinal disease \citep{aptos2019, rfmid2020} \\
    VTGAN \citep{kamran2021vtgan}                                             & 2021  & DFI                                                  & retinal                                                & retinal disease \citep{hajeb2012diabetic} \\
    MVT-based framework \citep{aladhadh2022effective}                         & 2022  & Dermoscopy                                           & skin                                                   & pigmented skin lesion \citep{tschandl2018ham10000} \\
    O-Net \citep{wang2022net}                                                 & 2022  & Dermoscopy                                           & skin                                                   & melanoma \citep{codella2018skin} \\     
    transformer \citep{hosain2022gastrointestinal}                            & 2022  & Endoscopy                                            & gastrointestinal                                       & gastrointestinal disease \citep{wceccd} \\
    transformer \citep{tamhane2022colonoscopy}                                & 2022  & Endoscopy                                            & colon                                                  & - \\
    transformer \citep{gheflati2022vision}                                    & 2022  & US                                                   & breast                                                 & breast disease \citep{yap2017automated, al2020dataset} \\
    multi-scale feature fusion transformer \citep{li2022crossa}               & 2022  & US                                                   & breast                                                 & - \\
    Advit \citep{xing2022advit}                                               & 2022  & PET                                                  & brain                                                  & Alzheimer’s Disease \citep{xing2021dynamic} \\
    multi-model transformer \citep{qayyum2021efficient}                       & 2021  & Camera                                               & toe                                                    & toe disease \citep{yap2021analysis} \\
    ViT-P \citep{wang2021vit}                                                 & 2021  & OCT                                                  & genitourinary                                          & genitourinary syndrome \citep{kermany2018identifying} \\
    SSBTN \citep{gong2022self}                                                & 2022  & X-ray, Microscope                                    & breast, small intestine                                & breast cancer \citep{moreira2012inbreast}, Crohn's disease \citep{vallee2020crohnipi} \\
    symmetric dual transformer \citep{al2022covid}                            & 2022  & X-ray, CT                                            & chest                                                  & lung disease \citep{wang2020covid, soares2020sars} \\
    grouped bottleneck transformer \citep{gao2022transformer}                 & 2022  & CT, MRI, Microscope                                  & tooth, abdomen, chest, brain, synapse                  & \citep{yang2021medmnist, yang2023medmnist} \\
    FPViT \citep{liu2022feature}                                              & 2022  & Microscope, X-ray, Dermoscopy, US, CT, DFI           & colon, chest, skin, chest, breast, abdomen, retinal    & \citep{yang2021medmnist} \\
    SEViT \citep{almalik2022self}                                             & 2022  & X-ray, DFI                                           & chest, retinal                                         & retinal disease \citep{aptos2019}, \citep{rahman2020reliable} \\
\bottomrule
\end{tabular*}
}
\label{tab1}
\end{table}

\textbf{X-ray}. Many of the researchers use existing transformer models to classify medical X-ray images \citep{ukwuoma2022automated, le2022covid, krishnan2021vision, behrendt2022data, ayana2023vision, jalalifar2022data, chen2022transformers}. The implemented transformer models include the transformer, the swin transformer, and the DeiT \citep{touvron2021training}. Besides using existing transformer models, some researcher aims at combining transformer with other CNN models. Within these, several papers tandems the CNN and transformer. For instance, Duong and co-workers \citep{duong2021detection} constructed a ViT-Eff method, where input images are fed into EfficientNet \citep{tan2019efficientnet} and the extracted feature maps are then projected into the transformer, followed by the developed classification head. Similar works include the PneuNet proposed by Wang et al. \citep{wang2023pneunet}, in which ResNet and transformer are tandem, and the Chest L-Transformer \citep{gu2022chest} that tandems ResNeXt \citep{xie2017aggregated} and transformer. Jalalifar et al. \citep{jalalifar2022data} built their method on a DeiT structure and let it benefits from the teacher-student scheme. The DenseNet \citep{huang2017densely} is set as the teacher while the adapted transformer is chosen as the student. Leveraging existing transformer models or combining the transformer with existing CNN models is intuitive and effective. However, the investigation of the model architecture is lacking, and potential performance improvement may be realized with an in-depth structure design.

Instead of combining existing models directly, several researchers build their models from scratch. Jiang et al. \citep{jiang2022mxt} designed an MXT method consisting of five stages. The first four stages are composed of several downsample spatial reduction transformer blocks and a multi-layer overlap patch embedding block. The last stage is composed of two class token transformer blocks and a multi-label attention block. Qi and co-workers \citep{qi2022multi} proposed a multi-feature fusion transformer where the cross-attention mechanism is deployed to learn information from both original images and corresponding enhanced local phase images. Jiang and Chen \citep{jiang2022multisemantic} developed an MP-ViT model, where images are fed to the patch fuser after enhancement and layer normalization. The obtained fusion features are then trained together with smoothed labels to obtain final prediction results. Park et al. \citep{park2021federated} developed a federated split transformer with the FESTA learning process. In FESTA, the server first initializes the weights of the transformer as well as task-specific heads and tails for each task. Then, it distributes the heads and tails weights to each client. For each round, each client (e.g., hospital) carries out the forward propagation on their head and conveys the intermediate feature to the server. Finally, the server aggregates and averages the weights of local heads and tails and distributes the updated global weights back to the clients.

\textbf{Microscope}. Some researchers utilize the existing transformer models to classify medical microscope images \citep{ikromjanov2022whole, zeid2021multiclass}. For instance, Zeid and co-workers \citep{zeid2021multiclass} proposed to use the transformer and the compact convolutional transformer \citep{hassani2021escaping}, in which the convolutional tokenizer is implemented instead of the patch-based tokenizer. The convolutional tokenization is composed of a convolutional layer, a ReLU activation function, followed by max pooling and reshaping, and can benefit the model from the inductive bias. Similar works include the explainable transformer-based model \citep{islam2022explainable}, in which the compact convolutional transformer and a gradient-weighted class activation map technique are implemented to show the attention paid to different parts by generating a heatmap. Some works made modifications based on the existing transformer model, such as the ensembled swin transformer proposed in 2022 \citep{tummala2022breast}. The ensembled swin transformer averages the predicted vectors of all individual ones. Tandeming the CNN and transformer is also commonly used. For instance, the IMGL-VTNet \citep{barmpoutis2022multi} tandems the ResNet and the deformable transformer encoder. The deformable transformer encoder is composed of a multi-scale deformable attention module and a feed-forward network. In the multi-scale deformable attention module, the multi-scale deformable attention function is leveraged to produce the feature map via weighted average. Zhang et al. \citep{zhang2022attention} proposed an AMIL-Trans network composed of two stages. In the first stage, features are captured using ResNet as well as the efficient channel attention module \citep{wang2020eca}. In the second stage, the transformer encoder for discriminant instance features takes the features as the input and outputs the prediction.

Instead of using the existing models or making minor modifications based on them, the remaining works constructed the novel models more deeply. Gul et al. \citep{gul2022histopathological} implemented a Self-ViT-MIL method, in which the transformer is first trained in a self-supervised manner using the DINO training approach. The multiple-instance learning aggregator is then trained with frozen transformer weights. Wang and co-workers \citep{wang2021transpath} developed a TransPath model consisting of a CNN encoder, a transformer encoder, and a token-aggregating and excitation module. The proposed self-supervised model is trained using the BYOL. Duan et al. \citep{duan2022fourier} constructed a Fourier ViT model consisting of two branches. The one branch is composed of two transformer encoders and their output information is exchanged with cross attention. Another branch normalizes the tokens and performs the 2D discrete Fourier transform. Elementwise multiplication is then implemented followed by the 2D inverse Fourier transform. The outputs of two encoders and the Fourier branch are concatenated before passing the fully connected layer. Lv and co-workers \citep{lv2022joint} constructed a RAMST, which can be further divided into the region-level RAMST and the WSI-level RAMST. Both region-level and WSI-level RAMST are composed of CNN and transformer while the WSI-level RAMST consists of an additional CNN branch. A novel feature weight uniform sampling method is also developed and implemented in both RAMSTs for patch subset sampling to preserve representative region features. In 2023, the \citep{wang2023cwc} CWC-Transformer is proposed by Wang et al. to solve the problem of feature extraction and spatial information loss effectively. In the compression stage, a feature compression method is implemented to extract discriminative features and reduce data bias. During the learning phase, the strengths of CNN and the transformer are extended to enhance the interrelationship between local and global information.

\textbf{CT}. A majority of researchers utilize developed transformer models to classify medical CT images \citep{gai2022using, sufian2022pre, li2021medical, salvi2022vision, sahoo2022vision, wu2021vision}. A typical example is the medical diagnostic platform developed by Li et al. \citep{li2021medical}. The platform is based on the transformer and can gain more medical information from the traditional image recognition model by distilling technology. There are a few works that contain novel transformer-based models. Xiong et al. \citep{xiong2022pulmonary} developed a multi-view convolutional transformer composed of four stages, which are view generation, visual backbone, feature decorrelation, and classifier. The visual backbone introduces the non-local self-attention into the last layer of ResNet, and the feature decorrelation learns a set of sample weights for eliminating the dependence between features. Mei \citep{mei2022marrying} married CNN and the transformer and constructed the DenseTransformer. The CNN and transformer are combined in three ways, including CNN and transformer in parallel, transformer in front of CNN in series, and CNN in front of transformer in series. Huang and co-workers \citep{huang2022transformer} proposed a transformer-based factorization encoder consisting of two transformer encoders. The former encoder enables the intra-slice interaction via encoding feature maps from the same slice, and the latter encoder investigates the inter-slice interaction via encoding feature maps from different slices. The multi-granularity dilated transformer \citep{wu2022multi} developed in 2022 leveraging the local focus scheme for guiding the deformable dilated transformer. The local focus scheme aims at discriminative local features more via modeling channel-wise grouped topology, and the deformable dilated transformer incorporates diverse contextual information.

\textbf{MRI}. Salanitri et al. proposed to leverage the transformer to diagnose intraductal papillary mucosal neoplasms \citep{salanitri2022neural}. Dai and co-workers \citep{dai2021transmed} developed a TransMed method by connecting the ResNet and the transformer in series. The double-scale generative adversarial network (GAN) method proposed by Hu et al. \citep{hu2021cross} is composed of a generator and two discriminators. The local CNN-based discriminator guides the generator to capture structural representation with inductive bias, while the transformer-based global discriminator directs the generator to extract comprehensive features via leveraging long-range dependencies. The MEST framework \citep{liu2022mest} developed in 2022 uses pre-trained VGGNet \citep{simonyan2014very} and attention mechanism to learn multi-plane dynamic images. Time-series information is used to construct dynamic functional connection images. Spatial-temporal connectivity transformer is utilized to solve spatiotemporal redundancy and dependencies, and ensemble learning is also employed to integrate multimodality data.

\textbf{DFI}. Yu and co-workers \citep{yu2021mil} developed a MIL-VT network. The MIL-VT uses the transformer as the backbone and a multiple-instance learning head is proposed to exploit the feature representations captured by individual patches better. The cross-entropy loss between the multilayer perceptron head and the label and between the multiple-instance learning head and label are computed. The VTGAN model \citep{kamran2021vtgan} constructed by Kamran et al. is composed of a coarse and a fine generator as well as two transformers as discriminators. The generators synthesize images according to the input and synthesized images are fed to the transformer encodes for classification. The two encoders also determine whether the synthesized images are from input or artificially generated.

\begin{figure*}
	\centering
		\includegraphics[width=0.82\textwidth]{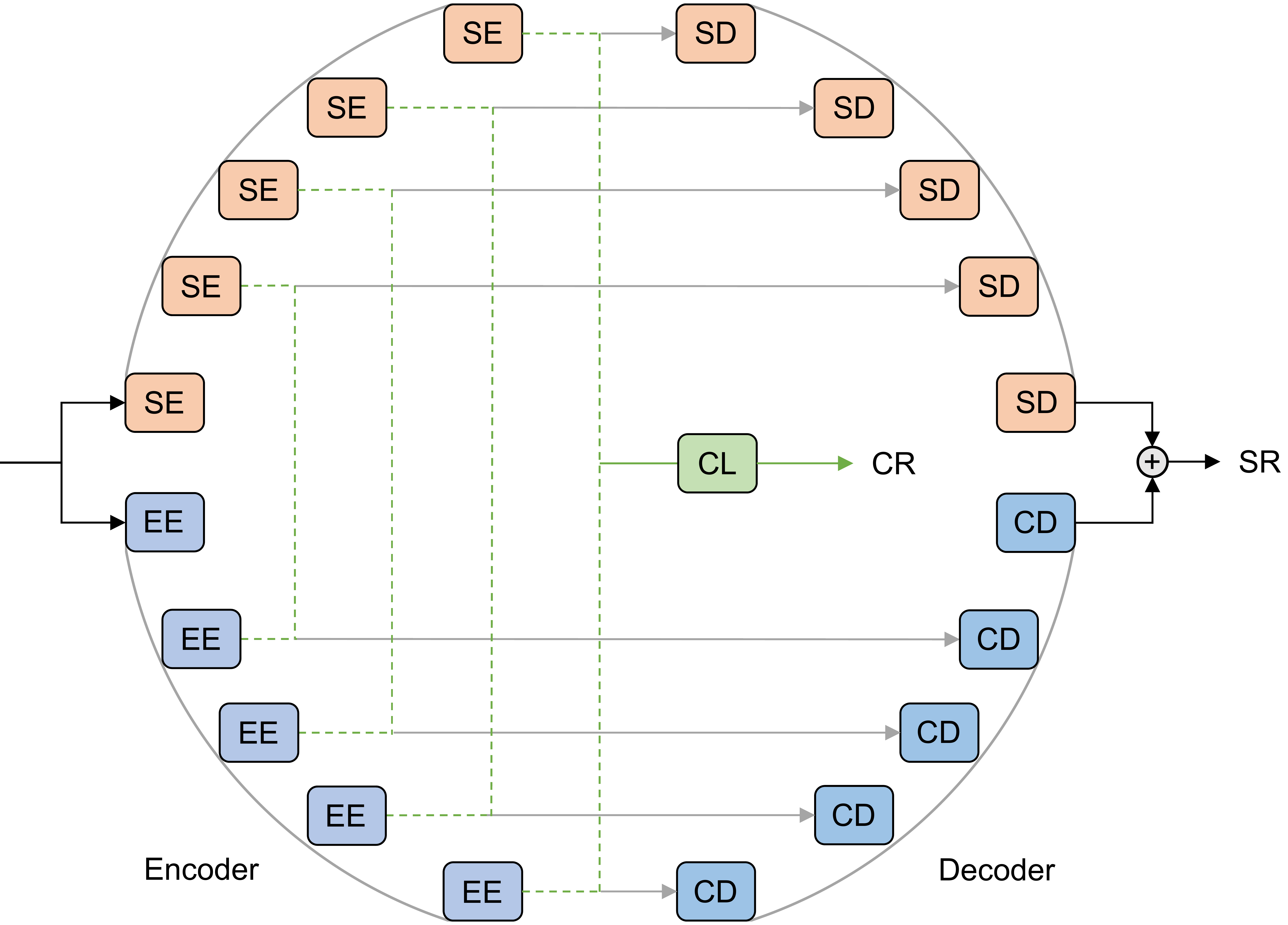}
	  \caption{Structure of the O-Net. O-Net has a novel architecture design and serves as a universal model for both classification and segmentation. EE, SE, CD, and SD represent the EfficientNet encoder block, swin transformer encoder block, CNN decoder block, and swin transformer decoder block, respectively. The gray solid lines show the skip connection. The green dotted lines depict fusion and unification. CL means the classification layer. CR represents the classification result and the SR illustrates the segmentation result.}
	  \label{fig9}
\end{figure*}

\textbf{Dermoscopy}. In 2022, Aladhadh et al. \citep{aladhadh2022effective} developed an MVT-based framework based on the transformer. Different data augmentation methods, including image flip, scaling, rotation, and contrast. Wang and co-workers designed a novel architecture O-Net, which serves as a universal model for both classification and segmentation. We show the complicated structure of the O-Net in \Cref{fig9}. The O-Net has a circle shape and is composed of four core blocks: EfficientNet encoder block, swin transformer encoder block, CNN decoder block, and swin transformer decoder block. Several EfficientNet encoder blocks and swin transformer encoder blocks composed the encoder, while the remaining two blocks compose the decoder. The encoder and decoder are connected by several skip connections.

\textbf{Endoscopy}. Hosain et al. \citep{hosain2022gastrointestinal} proposed to compare the classification performance between the transformer and CNN. Tamhane and co-workers \citep{tamhane2022colonoscopy} developed a landmark detection pipeline composed of three stages, including preprocessing, feature extracting, and classifying. The features are extracted using the transformer. CNN models including ResNet and ConvNeXt \citep{liu2022convnet} are also implemented for performance comparison.

\textbf{US}. Gheflati and Rivaz \citep{gheflati2022vision} utilize the transformer and compare its performance with several CNN models. Li et al. \citep{li2022crossa} developed a multi-scale feature fusion transformer by combining CNN and transformer. Short-distance feature interaction block is designed for the two feature maps within the CNN block, while a long-distance feature interaction block is developed for the feature maps between stages. Cross-SE block is introduced in the transformer block, which is mainly composed of the global average pooling layer and fully connected layer.

\textbf{PET}. Xing and co-workers \citep{xing2022advit} proposed an Advit model composed of two branches processing different modalities of PET. For each branch, a 3D-to-2D operation is implemented to project the 3D PET images into 2D fusion images using the proposed CNN module. The fused 2D images are then forwarded to the transformer. The output of two transformers is finally concatenated.

\textbf{Camera}. Qayyum et al. \citep{qayyum2021efficient} developed a multi-model transformer consisting of two separate pre-trained transformers. The outputs of both transformers are concatenated using pair-wise feature concatenation. The pair-wise feature concatenation is composed of two branches. In the first branch, the output of the second transformer is concatenated after the output of the first transformer. Regarding the second branch, the sequence is inverted. Outputs of two branches are then concatenated, in which the output of the first branch is placed in the front.

\textbf{OCT}. Wang and co-workers \citep{wang2021vit} developed a ViT-P architecture, consisting of a slim model and several transformer encoders in series. The model is mainly composed of four stages, in which each stage consists of the multi-branch convolutional and channel attention mechanism \citep{hu2018squeeze}. Besides, DCGAN \citep{radford2015unsupervised} and proposed B-DCGAN are implemented to perform data augmentation. Though their method shows promising results, we have to point out that the images generated by GAN are not quantitatively evaluated using metrics like IS \citep{salimans2016improved} and FID \citep{heusel2017gans} thus causing the performance difficult to represent intuitively.

\textbf{Multiple}. In 2022, Gong et al. \citep{gong2022self} developed an SSBTN composed of three modules, which are the pretext channel module, the transformer-based transfer module, and the downstream channel module. The transformer-based transfer module uses bi-channel transformer encoders and the loss between two channels is calculated. Rahhal and co-workers \citep{al2022covid} proposed a symmetric dual transformer consisting of two transformers. Original images are fed to one of the transformers with a class classifier while augmented images are sent to another transformer with a distill classifier. The outputs of the two transformers are then passed through a weighted fusion layer. Gao et al. \citep{gao2022transformer} built a grouped bottleneck transformer. The grouped bottleneck transformer block is composed of two branches, consisting of convolution operation only and both convolution operation and multi-head self-attention mechanism, respectively. The FPViT network \citep{liu2022feature} developed by Liu et al. extracts feature from three different layers from ResNet and input them into the transformer heads at different scales. The activation vectors are obtained through three transformers and the ResNet head, and prediction results are made through concatenated vectors. Almalik and co-workers \citep{almalik2022self} proposed a SEViT model, which trains separate multilayer perceptron layers using extracted patch tokens. A self-ensemble of different multilayer perceptron layers, together with the transformer classifier, enhances the robustness of the transformer. Besides, the consistency between ensemble predictions is leveraged for detecting adversarial samples.

\subsection{Segmentation}
\label{4.2}

Segmentation-related works are also grouped by different modalities in the sequence of MRI, CT, endoscopy, X-ray, US, microscope, DFI, camera, and dermoscopy. For papers containing more than one modality, we place them into the "multiple" category following the dermoscopy. The transformer-based segmentation works are organized in \Cref{tab2},  \Cref{tab3}, and \Cref{tab4} according to different modalities, respectively. Most of the segmentation works combine the transformer with the U-Net \citep{ronneberger2015u} or its variants like TransUNet \citep{chen2021transunet}. The U-Net is composed of a contracting path (encoder) following the structure of CNN, and an expansive path (decoder). The bottleneck layer is implemented between the encoder and decoder. The encoder is composed of several CNN blocks with the ReLU activation function. The output of each CNN block passes the max pooling for downsampling. With the downsampling, the number of feature channels is doubled. As for the decoder, it is composed of several upsampling, which halves the passed feature channels and several CNN blocks with the ReLU activation function. The features in the CNN blocks are concatenated with the feature obtained from the encoder at different scales. Specifically, the features obtained from a certain encoder scale are first cropped because border pixels are lost during convolution operation. The cropped features are then concatenated along the channel dimension.

\begin{table}
\centering
\caption{Transformer-based segmentation applications for CT and MRI modalities.}
\resizebox{\linewidth}{!}{
\begin{tabular*}{648pt}{ m{6.8cm}<{\centering} c m{3cm}<{\centering} m{2cm}<{\centering} m{8.2cm}<{\centering} }
\toprule
    Method                                                                        & Year  & Modality & Object                 & Dataset \\
\midrule
    UTNet \citep{gao2021utnet}                                                    & 2021  & MRI      & cardiac                & \citep{campello2021multi} \\
    MRA-TUNet \citep{chen2022multiresolution}                                     & 2022  & MRI      & cardiac                & cardiac disease \citep{bernard2018deep, xiong2021global} \\ 
    TransConver \citep{liang2022transconver}                                      & 2022  & MRI      & brain                  & brain tumor \citep{bakas2017advancing, bakas2018identifying, menze2014multimodal} \\
    UTransNet \citep{feng2022utransnet}                                           & 2022  & MRI      & brain                  & stroke \citep{atlasmri} \\
    TransBTS \citep{wang2021transbts}                                             & 2021  & MRI      & brain                  & brain tumor \citep{bakas2017advancing, bakas2018identifying, menze2014multimodal} \\
    METrans \citep{wang2022metrans}                                               & 2022  & MRI      & brain                  & stroke \citep{liew2018large, maier2017isles, islesischemic} \\ 
    SwinBTS \citep{jiang2022swinbts}                                              & 2022  & MRI      & brain                  & brain tumor \citep{bakas2017advancing, menze2014multimodal, bratstcgagbm} \\
    BTSwin-Unet \citep{liang2022btswin}                                           & 2022  & MRI      & brain                  & brain tumor \citep{bakas2017advancing, bakas2018identifying} \\  
    AST-Net \citep{wang2022ast}                                                   & 2022  & MRI      & brain                  & brain tumor \citep{menze2014multimodal} \\
    BiTr-Unet \citep{jia2022bitr}                                                 & 2022  & MRI      & brain                  & brain tumor \citep{menze2014multimodal} \\
    Swin UNETR \citep{hatamizadeh2022swin}                                        & 2022  & MRI      & brain                  & brain tumor \citep{menze2014multimodal} \\
    CST-UNET \citep{zhu20223d}                                                    & 2022  & MRI      & brain                  & brain tumor \citep{menze2014multimodal} \\
    VT-UNet \citep{peiris2022robust}                                              & 2022  & MRI      & brain                  & brain tumor \citep{menze2014multimodal} \\
    CSU-Net \citep{chen2022csu}                                                   & 2022  & MRI      & brain                  & brain tumor \citep{menze2014multimodal} \\     
    OSTransnet \citep{liu2022auxiliary}                                           & 2022  & MRI      & bone                   & osteosarcoma \citep{wu2022intelligent} \\
    3D PSwinBTS \citep{liang20223d}                                               & 2022  & MRI      & brain                  & brain tumor \citep{menze2014multimodal, antonelli2022medical} \\
    TSEUnet \citep{chen2022tseunet}                                               & 2022  & MRI      & brain                  & brain tumor \citep{menze2014multimodal} \\
    RMTF-Net \citep{gai2022rmtf}                                                  & 2022  & MRI      & brain                  & brain tumor \citep{brainmri, menze2014multimodal} \\   
    AMTNet \citep{zheng2022automated}                                             & 2023  & MRI      & prostate, brain        & brain tumor \citep{menze2014multimodal}, \citep{simpson2019large} \\
    transformer-based GAN \citep{huang2022transformera}                           & 2022  & MRI      & brain                  & brain tumor \citep{menze2014multimodal} \\
    DUconViT \citep{ling2022intelligent}                                          & 2022  & MRI      & bone                   & - \\
    CTCL \citep{li2022collaborative}                                              & 2022  & MRI      & cardiac                & cardiac disease \citep{bernard2018deep} \\
    symmetrical supervision transformer \citep{niu2022symmetrical}                & 2022  & MRI      & abdomen, cardiac       & \citep{kavur2021chaos, zhuang2018multivariate} \\
    transformer-enhanced U-Net \citep{gao2021consistency}                         & 2021  & MRI      & cardiac                & \citep{multiandmulti} \\
    TransUNet-based \citep{reyes2022gabor}                                        & 2022  & MRI      & brain, cardiac         & stroke\citep{islesischemic} \\
    dual-teacher \citep{xiao2022efficient}                                        & 2022  & MRI      & cardiac                & cardiac disease \citep{bernard2018deep} \\
    mmFormer \citep{zhang2022mmformer}                                            & 2022  & MRI      & brain                  & brain tumor \citep{menze2014multimodal} \\ 
    NVTrans-UNet \citep{li2023nvtrans}                                            & 2023  & MRI      & cardiac                & \citep{li2022myops} \\
    3D transformer \citep{karimi2022medical}                                      & 2022  & MRI      & brain                  & Alzheimer's \citep{mrihippo} \\
    iSegFormer \citep{liu2022isegformer}                                          & 2022  & MRI      & cartilage              & \citep{ambellan2019automated} \\
    transformer-based region-edge aggregation network \citep{chen2023transformer} & 2022  & MRI      & cardiac, knee          & cardiac disease \citep{xiong2021global} \\   
    CESS-ViT \citep{wang2022computationally}                                      & 2022  & MRI      & cardiac                & cardiac disease \citep{bernard2018deep} \\
    uncertainty-aware transformer \citep{wang2022uncertainty}                     & 2022  & MRI      & cardiac                & cardiac disease \citep{bernard2018deep} \\
    HybridCTrm \citep{sun2021hybridctrm}                                          & 2021  & MRI      & brain                  & \citep{mendrik2015mrbrains}, neurodevelopmental disorders \citep{wang2019benchmark} \\
    statistical features-based \citep{xu2022medical}                              & 2022  & MRI      & brain, cardiac         & \citep{antonelli2022medical}, brain tumor \citep{menze2014multimodal, bakas2018identifying, bakas2017advancing} \\
    feature fusion-based \citep{zhu2023brain}                                     & 2023  & MRI      & brain                  & brain tumor \citep{menze2014multimodal, bakas2018identifying, bakas2017advancing} \\
    UNTER \citep{mojtahedi2022towards}                                            & 2022  & CT       & liver                  & liver tumor \citep{bilic2019liver, simpson2017computed} \\
    CoTr \citep{xie2021cotr}                                                      & 2021  & CT       & abdomen                & colorectal cancer, ventral hernia \citep{synapseabdomen} \\    
    ITUnet \citep{kan2022itunet}                                                  & 2022  & CT       & head, neck             & - \\
    TFCNs \citep{li2022tfcns}                                                     & 2022  & CT       & abdomen, chest         & colorectal cancer, ventral hernia \citep{synapseabdomen}, lung disease \citep{covidcxr, covidctscan} \\
    TSE DeepLab \citep{yang2023tse}                                               & 2022  & CT       & sinus, patellar        & - \\
    transformer-UNet \citep{guo2021transformer}                                   & 2021  & CT       & lung                   & \citep{simpson2019large} \\    
    AFTer-UNet \citep{yan2022after}                                               & 2022  & CT       & abdomen, chest         & colorectal cancer, ventral hernia \citep{synapseabdomen}, organs at risk \citep{chen2021deep, lambert2020segthor} \\ 
    HT-Net \citep{ma2022ht}                                                       & 2022  & CT       & lung, kidney, bladder  & kidney tumor \citep{heller2019kits19}, lung lesion \citep{lungct}, bladder cancer \citep{bladderct} \\   
    UCATR \citep{luo2021ucatr}                                                    & 2021  & CT       & brain                  & - \\   
    MMViT-Seg \citep{yang2023mmvit}                                               & 2023  & CT       & lung                   & lung disease \citep{fan2020inf, covidmedicalseg} \\
    CCAT-net \citep{liu2022ccat}                                                  & 2022  & CT       & chest                  & lung disease \citep{morozov2020mosmeddata} \\
    CAC-EMVT \citep{ning2021cac}                                                  & 2021  & CT       & chest                  & - \\
    MSHT \citep{wang2021multi}                                                    & 2021  & CT       & liver, kidney          & kidney tumor \citep{heller2019kits19}, liver tumor \citep{bilic2019liver} \\    
    RCSHT \citep{ning2022hybrid}                                                  & 2022  & CT       & chest                  & - \\
    design-flexible transformer \citep{you2022more}                               & 2022  & CT       & liver, spine           & liver tumor \citep{antonelli2022medical}, \citep{sekuboyina2021verse} \\
    MAPTransNet \citep{dao2022survival}                                           & 2022  & CT       & lung                   & lung tumor \citep{aerts2014decoding, clark2013cancer} \\
    CTUNet \citep{chen2022ctunet}                                                 & 2022  & CT       & pancreas               & \citep{roth2015deep} \\
\bottomrule
\end{tabular*}
}
\label{tab2}
\end{table}

\textbf{MRI}. The U-Net and its variants are widely used to segment medical MRI images. Within these works, most of the works aim at modifying the encoder, decoder, or the bottleneck layer \citep{gao2021utnet, chen2022multiresolution, liang2022transconver, feng2022utransnet, wang2021transbts, wang2022metrans, jiang2022swinbts, liang2022btswin, wang2022ast, jia2022bitr, hatamizadeh2022swin, zhu20223d, peiris2022robust, chen2022csu, liu2022auxiliary, liang20223d, chen2022tseunet, gai2022rmtf, zheng2022automated, huang2022transformera}. Many works utilize several U-Nets when constructing their model \citep{ling2022intelligent, li2022collaborative, niu2022symmetrical, gao2021consistency, reyes2022gabor, xiao2022efficient}. For instance, Xiao et al. \citep{xiao2022efficient} developed a semi-supervised dual-teacher architecture, which uses simultaneous dual-teacher to guide the student. The two teachers use U-Net and Swin-UNet \citep{cao2021swin} as the backbone and the student uses the U-Net as the backbone. The U-Net teacher and U-Net student, and two teachers inside are screened for uncertainty assessment during training. Beyond single modal, there are also several works aiming at multimodal tasks \citep{zhang2022mmformer, li2023nvtrans}, such as the multimodal network NVTrans-UNet \citep{li2023nvtrans} developed by Li et al. The input of the NVTrans-UNet is composed of three main parts, including the encoder module, bottleneck layer, and decoder module. The NVTrans-UNet utilizes the neighborhood transformer to localize the receptive field of each token to its nearest neighboring pixel. The multi-modal gated fusion strategy is implemented to adjust the contribution of feature mapping from each modal. Atrous spatial pyramid pooling is also used in the bottleneck layer for expanding the receptive field, reducing parameters, and enhancing extraction ability.

Besides the U-Net-based work, there are several models developed in other ways. Several works made minor modifications based on the existing model or connect two models in series. For example, Karimi et al. \citep{karimi2022medical} developed a 3D transformer, in which the residual connection in the transformer encoder block is removed. Liu and co-workers \citep{liu2022isegformer} proposed an iSegFormer, where the swin transformer and lightweight multilayer perceptron decoder are combined in series. Some researchers design their models from scratch \citep{chen2023transformer, wang2022computationally, wang2022uncertainty, sun2021hybridctrm, xu2022medical, zhu2023brain}. For instance, Chen et al. \citep{chen2023transformer} proposed a transformer-based region-edge aggregation network, where the multi-level region and edge features are aggregated by multiple transformer-based inference modules to form multi-level complementary features. These complementary features are utilized to guide the decoding of the corresponding level region and edge features. Sun and co-workers \citep{sun2021hybridctrm} developed a multimodal HybridCTrm network, which is composed of two paths. The first path takes the MRI-T1 and MRI-T2 images together, followed by the parallel CNN and transformer, while the second path takes the MRI-T1 and MRI-T2 images separately. In 2023, a novel method based on deep semantics and edge information fusion is developed \citep{zhu2023brain}. The proposed method is composed of a semantic segmentation module, an edge detection module, as well as a feature fusion module. The segmentation module utilizes the swin transformer as the backbone with shifted patch tokenization strategy. The CNN-based detection module consists of the proposed edge spatial attention block for feature enhancement. Semantic and edge features from two modules are fused by the feature fusion module.

\textbf{CT}. Instead of using existing models to segment CT images, such as the framework \citep{mojtahedi2022towards} using UNTER \citep{hatamizadeh2022unetr} as the backbone, major works use U-Net-based novel networks with different aspects of modifications \citep{xie2021cotr, kan2022itunet, li2022tfcns, yang2023tse, guo2021transformer, yan2022after, ma2022ht, luo2021ucatr, liu2022ccat, ning2021cac, wang2021multi, ning2022hybrid, you2022more, dao2022survival, chen2022ctunet, yang2023mmvit}. For instance, Xie et al. \citep{xie2021cotr} designed a framework CoTr with three parts: A CNN encoder, a DeTrans encoder, and a Decoder. The DeTrans encoder connects the CNN encoder and the decoder at different scales. The output of the CNN encoder is flattened before feeding to the DeTrans encoder and the output of the DeTrans encoder is reshaped and then send to the decoder. The DeTrans only pays attention to a small set of key positions thus the complexity is reduced largely. Kan and co-workers \citep{kan2022itunet} developed an ITUnet, in which the feature map of CNN and transformer are added in the downsampling stage. In the upsampling stage, the segmentation predictions for each feature map obtained by the up block are generated and utilized to calculate the loss. Li et al. \citep{li2022tfcns} constructed a TFCNs model, in which the encoder is constructed by introducing the transformer into the FC-DenseNet \citep{jegou2017one}. The RL-Transformer layer is added at the end of the encoder and the convolutional linear attention block is introduced in the skip connection to filter non-semantic features by including spatial and channel attention. In 2023, Yang and co-workers \citep{yang2023tse} developed a TSE DeepLab framework, which leverages atrous convolution in DeepLabv3 \citep{chen2017rethinking} as the backbone to extract features. The captured features are then converted into visual tokens and then fed to the transformer. Squeeze and excitation components are also introduced after the transformer for channel importance sorting.

\begin{table}
\centering
\caption{Transformer-based segmentation applications for endoscopy, X-ray, US, microscope, DFI, camera, and dermoscopy modalities.}
\resizebox{\linewidth}{!}{
\begin{tabular*}{600pt}{ m{6.8cm}<{\centering} c m{2.8cm}<{\centering} m{2.6cm}<{\centering} m{5.8cm}<{\centering} }
\toprule
    Method                                                            & Year  & Modality       & Object           & Dataset \\
\midrule  
    RANT \citep{pan2022rant}                                          & 2022  & Endoscopy      & throat           & \citep{laves2019dataset} \\    
    BiDFNet \citep{tang2022bidfnet}                                   & 2022  & Endoscopy	   & colon            & polyp \citep{jha2020kvasir, bernal2015wm, silva2014toward, tajbakhsh2015automated, vazquez2017benchmark} \\
    Patcher \citep{ou2022patcher}                                     & 2022  & Endoscopy	   & colon            & polyp \citep{jha2020kvasir} \\
    Polyp2Seg \citep{mandujano2022polyp2seg}                          & 2022  & Endoscopy      & colon            & polyp \citep{jha2020kvasir, bernal2015wm, silva2014toward, tajbakhsh2015automated, vazquez2017benchmark} \\
    TransHarDNet \citep{wang2022medicala}                             & 2022  & Endoscopy      & colon            & polyp \citep{jha2020kvasir, tajbakhsh2015automated, silva2014toward, vazquez2017benchmark, bernal2015wm} \\
    FCBFormer \citep{sanderson2022fcn}                                & 2022  & Endoscopy      & colon            & polyp \citep{jha2020kvasir, cvcclinicdb} \\
    U-Net \citep{saidnassim2021self}                                  & 2021  & X-ray          & breast           & - \\
    temporary transformer \citep{zhang2021temporary}                  & 2022  & X-ray          & catheter         & - \\   
    APSegmenter \citep{zhang2022spine}                                & 2022  & X-ray          & spine            & spinal curvature \citep{aasce} \\
    Chest L-Transformer \citep{gu2022chest}                           & 2022  & X-ray          & chest            & lung disease \citep{filice2020crowdsourcing} \\
    federated split transformer \citep{park2021federated}             & 2021  & X-ray          & lung             & lung disease \citep{siimacr} \\
    TransBridge \citep{deng2021transbridge}                           & 2021  & US             & cardiac          & cardiac disease \citep{ouyang2020video} \\    
    TFNet \citep{wang2022tfnet}                                       & 2022  & US             & breast, thyroid  & breast disease \citep{al2020dataset}, thyroid disorder \citep{pedraza2015open} \\
    CSwin-PNet \citep{yang2023cswin}                                  & 2023  & US             & breast           & breast disease \citep{yap2017automated, al2020dataset} \\
    RSTUnet-CR \citep{zhuang2022residual}                             & 2022  & US             & breast           & - \\
    dilated transformer \citep{shen2022dilated}                       & 2022  & US             & breast           & breast tumor \citep{zhang2022busis} \\ 
    Swin-PANet \citep{liao2022swin}                                   & 2022  & Microscope     & colon, cell      & colon cancer \citep{sirinukunwattana2017gland}, \citep{kumar2017dataset} \\
    multiple-instance transformer \citep{qian2022transformer}         & 2022  & Microscope     & colon            & colon cancer \citep{jia2017constrained} \\ 
    SMESwin Unet \citep{wang2022smeswin}                              & 2022  & Microscope     & colon, cell      & colon cancer \citep{sirinukunwattana2017gland}, \citep{kumar2017dataset}, \citep{koohbanani2020nuclick} \\
    PCAT-UNet \citep{chen2022pcat}                                    & 2022  & DFI            & retinal          & retinal disease \citep{staal2004ridge, owen2009measuring}, \citep{hoover2000locating} \\ 
    Polarformer \citep{feng2022polarformer}                           & 2022  & DFI            & retinal          & retinal disease \citep{orlando2020refuge, sivaswamy2014drishti, fumero2011rim} \\
    GT-DLA-dsHFF \citep{li2022global}                                 & 2022  & DFI            & retinal          & retinal disease \citep{staal2004ridge, hoover2000locating, cherukuri2019deep}, \citep{fraz2012ensemble} \\
    versatile transformer \citep{junayed2022transformer}              & 2022  & Camera         & skin             & - \\
    semi-supervised transformer \citep{alahmadi2022semi}              & 2022  & Dermoscopy     & skin             & melanoma \citep{codella2018skin, codella2019skin, mendoncca2013ph} \\
\bottomrule
\end{tabular*}
}
\label{tab3}
\end{table}

\textbf{Endoscopy}. Pan et al. \citep{pan2022rant} proposed a RANT framework, in which the transformer and CNN are combined in series. Features are cascaded using reverse attention and receptive field block module. The segmentation results are optimized using convolutional conditional random fields. Tang and co-workers \citep{tang2022bidfnet} proposed a bi-decoder BiDFNet works in both fine-to-coarse and coarse-to-fine ways. The BiDFNet is composed of an encoder based on PVTv2 \citep{wang2021pyramid} as well as two decoders connected in series. The adaptive fusion module and the residual connection module are implemented in the decoders, and the adaptive fusion module aggregates the features from different scales effectively. Ou et al. \citep{ou2022patcher} developed a Patcher method, where the encoder utilizes a cascade of Patcher blocks for expert features capture at different scales. The Patcher block first segment the input to large patches with overlapping contexts and then further divided them into small patches. The divided small patches are next fed to sequential transformer blocks for feature extraction and the large patches are finally reassembled. The mixture-of-experts-based decoder utilizes a gating network to filter a set of suitable expert features for the prediction. Mandujano and co-workers \citep{mandujano2022polyp2seg} constructed a Polyp2Seg network, which uses PVTv2 to extract a set of multi-scale features. The extracted multi-scale features are then compressed and fed into several feature aggregation modules. A multi-context attention module is implemented to characterize low-level polyp cues and the final predicted results are obtained by several auxiliary outputs. The TransHarDNet network \citep{wang2022medical} designed in 2022 combines the transformer and HarDNet blocks \citep{chao2019hardnet}. HarDNet Blocks are leveraged to extract spatial and depth information, while the transformer captures global semantic context information. Several cascaded partial decoders are implemented to fuse the feature maps and the skip connection with the receptive field block is implemented between the HarDNet blocks and partial decoders. Sanderson et al. \citep{sanderson2022fcn} designed an FCBFormer architecture consisting of two branches. The transformer branch extracts the most important features based on the transformer, while the fully convolutional branch is implemented as a supplementary. The output of the two branches is then concatenated and passes the prediction head.

\textbf{X-ray}. Saidnassim and co-workers \citep{saidnassim2021self} proposed to use the BYOL algorithm for U-Net-based breast image segmentation. Zhang et al. \citep{zhang2021temporary} proposed a temporary transformer network, which takes both the current and previous frames as the input to obtain temporary information. The current frame is fed into the CNN and transformer, while the previous frame is fed into the CNN only. In 2022, Zhang and co-workers \citep{zhang2022spine} developed an APSegmenter method in which the transformer-based Segmenter \citep{strudel2021segmenter} is utilized to obtain semantic segmentation results. The proposed adaptive post-processing module is utilized to optimize the results, which takes the vertebral block boundary in the adhesion region as the input and outputs the vertebral mass without adhesion. The Chest L-Transformer and federated split transformer discussed in \Cref{tab1} are also leveraged for medical image segmentation.

\begin{figure*}
	\centering
		\includegraphics[width=0.76\textwidth]{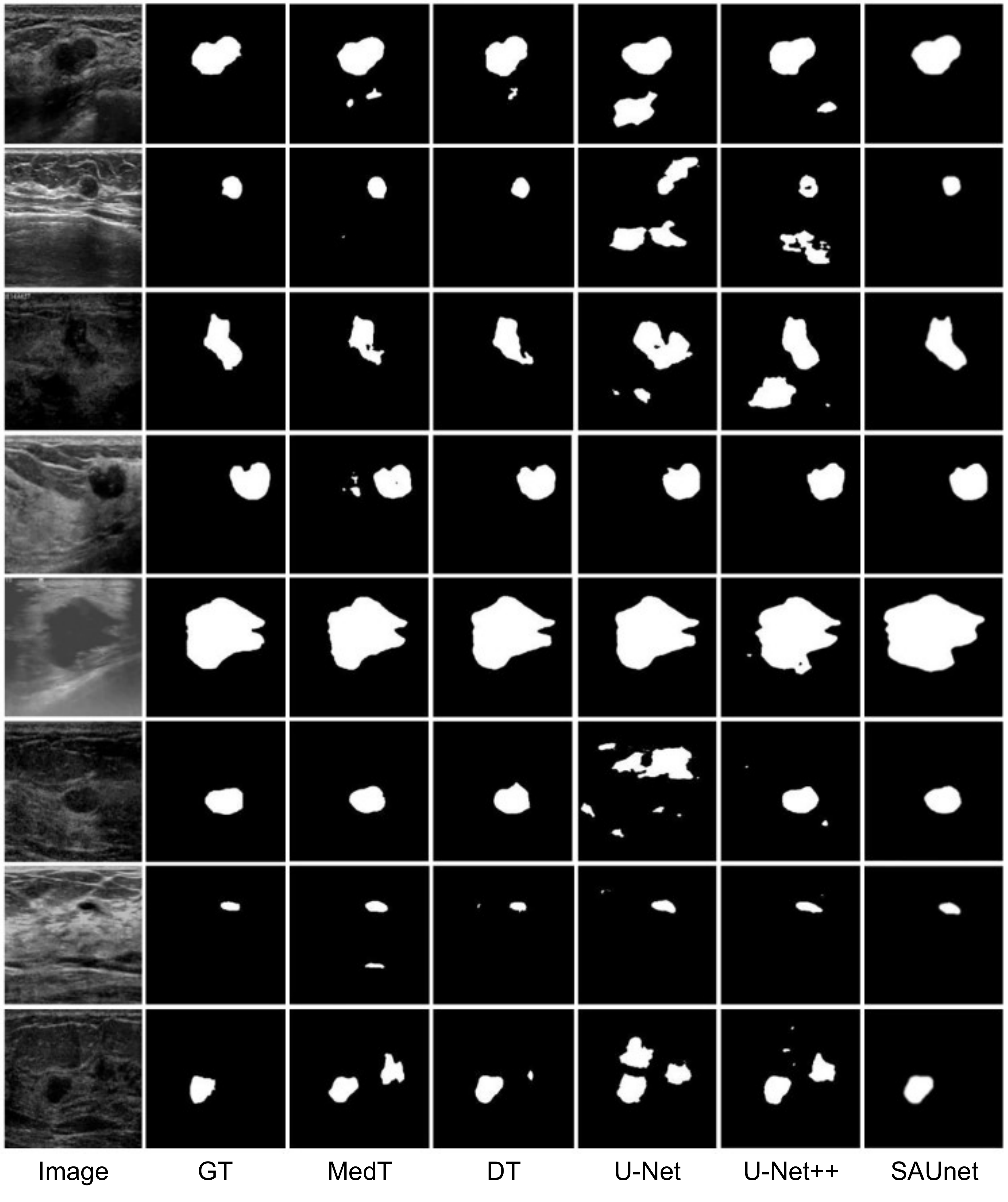}
	  \caption{Segmentation results on the BUSIS dataset \citep{zhang2022busis} using MedT \citep{valanarasu2021medical}, dilated transformer (DT) \citep{shen2022dilated}, U-Net \citep{ronneberger2015u}, U-Net++ \citep{zhou2018unet++}, and SAUnet \citep{vakanski2020attention}. GT represents the ground truth.}
	  \label{fig10}
\end{figure*}

\textbf{US}. Deng et al. \citep{deng2021transbridge} proposed a TransBridge model, in which both the encoder and decoder are based on CNN, while the transformer encoder is used to skip-connect them at different scales. Within the transformer encoder, an embedding layer is implemented by using shuffled group convolution and dense patch division. Wang and co-workers \citep{wang2022tfnet} constructed a TFNet model, where the channel attention mechanism is introduced for solving the channel modeling defect. A loss function based on KL distance is also proposed to modify the predicted results by calculating the variance between the results of the main classifier and the auxiliary classifier. In 2023, Yang et al. \citep{yang2023cswin} built a CSwin-PNet model. An interactive channel attention module using channel-wise attention and an SFF module is developed for feature region emphasize and feature supplementary during fusion, respectively. Besides, a boundary detection module is also utilized to extract the boundary information. Zhuang and co-workers \citep{zhuang2022residual} designed an RSTUnet-CR model consisting of a shared encoder, a segmentation decoder, and a consistency regularization decoder where long-distance dependence is established using the residual swin transformer block. The dilated transformer model proposed in 2022 \citep{shen2022dilated} uses a dilation convolution block to connect the encoder and decoder. The encoder contains the multi-head attention mechanism and the decoder is mainly composed of deconvolution. As shown in \Cref{fig10}, the dilated transformer performs better compared with other state-of-the-art methods. Among all methods, the dilated transformer and the SAUnet \citep{vakanski2020attention} perform better due to low false positives, meaning that the boundaries can be distinguished precisely. Within the two methods, the dilated transformer can capture information in more detail. Take the fourth row as an example, the dilated transformer can distinguish the invaginated part at the top tumor better. Though the dilated transformer outperforms other models, we also observed that it can sometimes produce unideal results. Several examples would be the first, seventh, and eighth rows, in which an isolated extra object is mistakenly created. This means the ability to differentiate textures with subtle differences still has a margin to be improved.

\textbf{Microscope}. In 2022, Liao et al. \citep{liao2022swin} proposed a Swin-PANet model following the coarse-to-fine as well as dual supervision strategy. The developed Swin-PANet consists of a prior attention network and a hybrid transformer network. The swin transformer-assisted prior attention network carries out intermediate supervision learning, while the hybrid transformer network with enhanced attention blocks implements direct learning. Besides, the skip connection is employed to connect the encoder and decoder of the hybrid transformer network. Qian and co-workers \citep{qian2022transformer} developed a multiple-instance transformer where the transformer is incorporated into the multiple-instance learning framework. The self-attention establishes the relationship among different instances. Deep supervision is implemented to overcome the annotation limitation existing in weakly-supervised methods. Wang et al. \citep{wang2022smeswin} developed a SMESwin Unet model based on their proposed MCCT. The MCCT is designed to fuse multi-scale semantic features and attention maps based on the channel-wise cross-fusion transformer \citep{wang2022uctransnet}. Superpixel is introduced by dividing the pixel-level feature into district-level and external attention is leveraged to introduce the correlations among all data samples.

\textbf{DFI}. Chen and co-workers \citep{chen2022pcat} designed a PCAT-UNet model containing two main components named patches convolution attention transformer block and feature grouping attention module. Both encoder and decoder are composed of several patches convolution attention transformer blocks and the outputs of the feature grouping attention modules are fed into the patches convolution attention transformer blocks. The segmentation map of each layer is predicted using the fused enhanced feature map. Feng et al. \citep{feng2022polarformer} proposed a Polarformer network mainly composed of the learnable polar transformation module as well as the CNN-transformer module. The polar transformation module carries out a differentiable log-polar transform, while the CNN-transformer module captures features and consolidates global attention. A segmentation head is implemented to output the confidence scores and transmute the predictions back to the Cartesian coordinate system. Li and co-workers \citep{li2022global} developed a GT-DLA-dsHFF model, in which a global transformer and dual local attention network are introduced for global information integration and local vessel information extraction, respectively. Besides, a deep-shallow hierarchical feature algorithm is used to fuse features.

\textbf{Camera}. The versatile transformer \citep{junayed2022transformer} developed by Junayed et al. is composed of the dual encoder, the feature versatile block, and efficient decoder architecture with skip connections. The dual encoder is based on CNN and transformer to extract features, and the feature versatile block is implemented to distribute and integrate obtained features between the encoder and decoder. A squeeze and excitation block component is also introduced in the decoder to capture channel-wise dependencies as well as the significant feature correlations.

\textbf{Dermoscopy}. Alahmadid and co-workers \citep{alahmadi2022semi} proposed a transformer consisting of a supervised stream and an unsupervised stream. The supervised stream combines CNN and transformer and the output features of CNN and transformer are fused. Specifically, the transformer output is reshaped into the same spatial dimension as the CNN, and then two features are concatenated. The fused features are then fed to the decoder module for semantic segmentation learning. The unsupervised stream is composed of a supplementary decoding head and utilizes the unsupervised technique for encoder module enrichment. A surrogate task is designed on top of the CNN and transformer representations.

\begin{table}
\centering
\caption{Transformer-based segmentation applications for multiple modalities.}
\resizebox{\linewidth}{!}{
\begin{tabular*}{700pt}{ m{4.8cm}<{\centering} c m{4.6cm}<{\centering} m{4cm}<{\centering} m{8cm}<{\centering} }
\toprule
    Method                                                        & Year  & Modality                                            & Object                          & Dataset \\
\midrule  
    MISSFormer \citep{huang2022missformer}                        & 2022  & CT, MRI, DFI                              & abdomen, cardiac, retinal       & colorectal cancer, ventral hernia \citep{synapseabdomen}, cardiac disease \citep{acdcmri}, retinal disease \citep{staal2004ridge} \\
    Dual encoder transformer-CNN \citep{hong2022dual}             & 2022  & CT, MRI                                             & abdomen, cardiac                & colorectal cancer, ventral hernia \citep{synapseabdomen}, cardiac disease \citep{acdcmri} \\   
    ConTrans \citep{lin2022contrans}                              & 2022  & Endoscopy, Dermoscopy, CT, Microscope               & cell, skin, chest, colon        & polyp \citep{jha2020kvasir, silva2014toward, vazquez2017benchmark, bernal2015wm}, melanoma \citep{codella2018skin, codella2019skin}, pigmented skin lesion \citep{tschandl2018ham10000}, lung disease \citep{covidct}, colon cancer \citep{sirinukunwattana2017gland}, \citep{caicedo2019nucleus}, cancer \citep{gamper2019pannuke} \\
    ScaleFormer \citep{huang2022scaleformer}                      & 2022  & CT, MRI, Microscope                                 & abdomen, cell, cardiac          & \citep{landman2015segmentation, kumar2017dataset}, cardiac disease \citep{bernard2018deep} \\
    EMSViT \citep{sagar2022emsvit}                                & 2022  & MRI, CT                                             & abdomen, brain                  & colorectal cancer, ventral hernia \citep{synapseabdomen}, brain tumor \citep{menze2014multimodal, bakas2018identifying} \\    
    TransCUNet \citep{jiang2022transcunet}                        & 2022  & Microscope, Endoscopy, Dermoscopy                   & colon, cell, skin               & colon cancer \citep{sirinukunwattana2017gland}, polyp \citep{bernal2015wm}, \citep{kumar2019multi, kumar2017dataset}, melanoma \citep{codella2018skin} \\
    CATS \citep{li2022cats}                                       & 2022  &	CT, MRI                                             & abdomen, brain, prostate 	      & brain tumor \citep{menze2014multimodal}, vestibular schwannomas \citep{moda}, \citep{decfive} \\
    D-former \citep{wu2022d}                                      & 2022  &	CT, MRI	                                            & abdomen, cardiac 	              & brain tumor \citep{menze2014multimodal}, cardiac disease \citep{bernard2018deep} \\
    APT-Net \citep{zhang2022apt}                                  & 2022  &	Dermoscopy, Endoscopy, Microscope                   & skin, colon                 	  & melanoma \citep{codella2018skin, mendoncca2013ph}, polyp \citep{jha2020kvasir, silva2014toward, bernal2015wm, tajbakhsh2015automated, vazquez2017benchmark}, colon cancer \citep{sirinukunwattana2017gland} \\
    TransNorm \citep{azad2022transnorm}                           & 2022  &	CT, Dermoscopy, Microscope                          & abdomen, skin, cell	          & colorectal cancer, ventral hernia \citep{synapseabdomen}, melanoma \citep{codella2018skin, codella2019skin, mendoncca2013ph}, myeloma \citep{gupta2018pcseg} \\
    SwinPA-Net \citep{du2022swinpa}                               & 2022  & Colonoscopy, Microscope, Camera	                    & colon, cell    	              & polyp \citep{jha2020kvasir, silva2014toward, bernal2015wm, tajbakhsh2015automated, vazquez2017benchmark}, \citep{caicedo2019nucleus} \\    
    GPA-TUNet \citep{li2022transformer}	                          & 2022  &	CT, MRI                                             & abdomen, cardiac	              & colorectal cancer, ventral hernia \citep{synapseabdomen}, cardiac disease \citep{bernard2018deep} \\
    ConvWin-UNet \citep{feng2023convwin}                          & 2023  & Microscope, CT                                      & kidney, abdomen                 & colorectal cancer, ventral hernia \citep{synapseabdomen}, \citep{hubmap} \\
    PCT \citep{zhang2023pct}                                      & 2023  & US, Microscope, Dermoscopy	                        & parotid, skin, cell             & \citep{kumar2017dataset}, melanoma \citep{codella2018skin} \\ 
    DS-TransUNet \citep{lin2022ds}                                & 2022  & Endoscopy, Dermoscopy, Microscope                   & colon, skin, cell               & colorectal cancer \citep{silva2014toward}, colon cancer \citep{sirinukunwattana2017gland}, polyp \citep{jha2020kvasir, bernal2015wm, tajbakhsh2015automated, vazquez2017benchmark}, \citep{caicedo2019nucleus}, melanoma \citep{codella2019skin} \\
    DSTUNet \citep{cai2022dstunet}                                & 2022  & MRI, CT                                             & abdomen, cardiac                & cardiac disease \citep{bernard2018deep}, colorectal cancer, ventral hernia \citep{synapseabdomen}, cardiac disease \citep{acdcmri} \\
    MT-UNet \citep{wang2022mixed}                                 & 2022  & CT, MRI                                             & abdomen, cardiac                & colorectal cancer, ventral hernia \citep{synapseabdomen}, cardiac disease \citep{acdcmri} \\
    ViTBIS \citep{sagar2021vitbis}                                & 2021  & CT, MRI                                             & abdomen, brain                  & brain tumor \citep{menze2014multimodal, bakas2018identifying}, colorectal cancer, ventral hernia \citep{synapseabdomen} \\      
    TDD-UNet \citep{huang2022tdd}                                 & 2022  & CT, X-ray                                           & lung                            & lung disease \citep{covidmedicalseg, qatacovid} \\
    SwinE-Net \citep{park2022swine}                               & 2022  & Endoscopy, MRI                                      & colon, brain                    & polyp \citep{jha2020kvasir, bernal2015wm, tajbakhsh2015automated, silva2014toward, vazquez2017benchmark} \\
    USegTransformer \citep{dhamija2022semantic}                   & 2022  & Dermoscopy, MRI, CT, Microscope                     & brain, lung, cell, skin, chest  & pigmented skin lesion \citep{tschandl2018ham10000}, lung lesion \citep{lungct}, brain tumor \citep{brainmri}, \citep{caicedo2019nucleus}, melanoma \citep{codella2019skin}, lung disease \citep{covidct} \\
    SegTransVAE \citep{pham2022segtransvae}                       & 2022  & CT, MRI                                             & kidney, brain                   & kidney tumor \citep{heller2019kits19}, brain tumor \citep{menze2014multimodal} \\
    MedT  \citep{valanarasu2021medical}                           & 2021  & US, Microscope                                      & brain, colon, cell              & intraventricular hemorrhage \citep{valanarasu2020learning, wang2018automatic}, colon cancer \citep{sirinukunwattana2017gland},  \citep{kumar2017dataset, kumar2019multi} \\
    medical transformer \citep{tang2023combined}                  & 2023  & MRI, US, Camera	                                    & prostate, cardiac, tongue       & \citep{tongeimage} \\
    CTC-Net \citep{yuan2023effective}                             & 2023  &	CT, MRI                                             & abdomen, cardiac	              & colorectal cancer, ventral hernia \citep{synapseabdomen}, cardiac disease \citep{bernard2018deep} \\
    ST-Unet \citep{zhang2023st}                                   & 2023  & Dermoscopy, CT                                      & skin, abdomen                   & colorectal cancer, ventral hernia \citep{synapseabdomen}, pigmented skin lesion \citep{tschandl2018ham10000}, melanoma \citep{codella2019skin} \\
    MS-TransUNet++ \citep{wang2022multiscale}                     & 2022  & MRI, CT                                             & prostate, liver                 & liver tumor \citep{bilic2019liver}, prostate cancer \citep{litjens2014evaluation} \\   
    O-Net \citep{wang2022net}                                     & 2022  & Dermoscopy, CT                                      & skin, abdomen                   & melanoma \citep{codella2018skin}, colorectal cancer, ventral hernia \citep{synapseabdomen} \\
    TMSS \citep{saeed2022tmss}                                    & 2022  & PET, CT                                             & head, neck                      & tumor \citep{hecktor2021} \\
    TransDeepLab \citep{azad2022transdeeplab}                     & 2022  & CT, Dermoscopy                                      & abdomen, skin                   & colorectal cancer, ventral hernia \citep{synapseabdomen}, melanoma \citep{codella2019skin, codella2018skin, mendoncca2013ph} \\
    transformer \citep{wang2023self}                              & 2023  & CT, MRI	                                            & abdomen	                      & colorectal cancer, ventral hernia \citep{synapseabdomen}, \citep{kavur2021chaos} \\
    ECT-NAS \citep{xu2021ect}                                     & 2021  & CT, MRI                                             & abdomen, cardiac                & cardiac disease \citep{bernard2018deep}, \citep{gibson2018multi}, \citep{kavur2021chaos} \\    
    SMIT \citep{jiang2022self}                                    & 2022  & CT, MRI                                             & abdomen                         & colorectal cancer, ventral hernia \citep{synapseabdomen} \\
    progressive sampling transformer \citep{jiang2022transformer} & 2022  & Microscope, Endoscopy                               & colon, cell	                  & colon cancer \citep{sirinukunwattana2017gland}; polyp \citep{bernal2015wm}, \citep{kumar2019multi, kumar2017dataset} \\
    X-Net \citep{li2021x}                                         & 2021  & Microscope, Endoscopy                               & colon, cell                     & \citep{caicedo2019nucleus, naylor2018segmentation}, polyp \citep{jha2020kvasir} \\
\bottomrule
\end{tabular*}
}
\label{tab4}
\end{table}

\begin{figure*}
	\centering
		\includegraphics[width=0.48\textwidth]{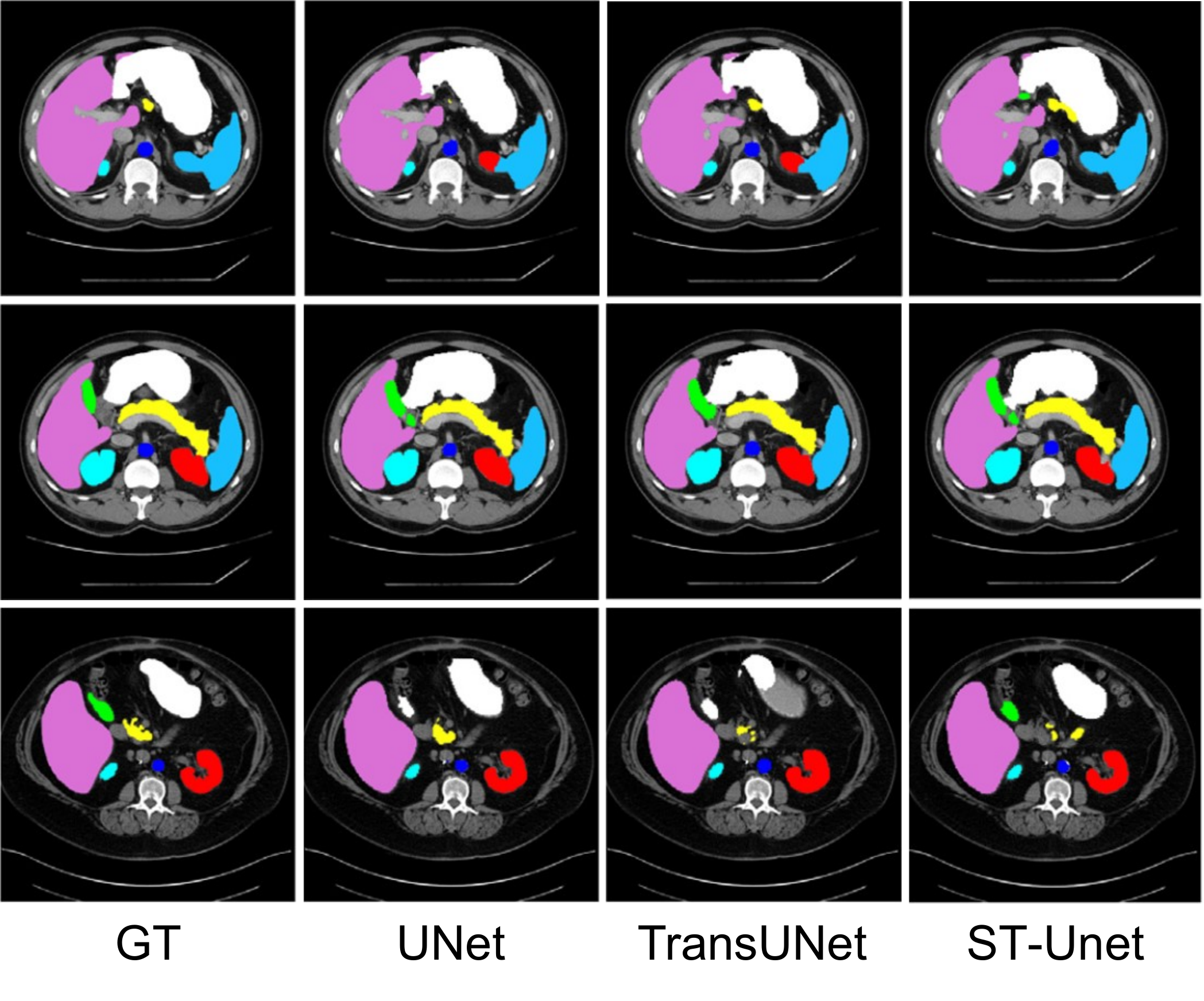}
	  \caption{Segmentation results on the Synapse dataset \citep{synapseabdomen} using UNet, TransUNet \citep{chen2021transunet}, and ST-Unet \citep{zhang2023st}.}
	  \label{fig11}
\end{figure*}

\begin{figure*}
	\centering
		\includegraphics[width=0.90\textwidth]{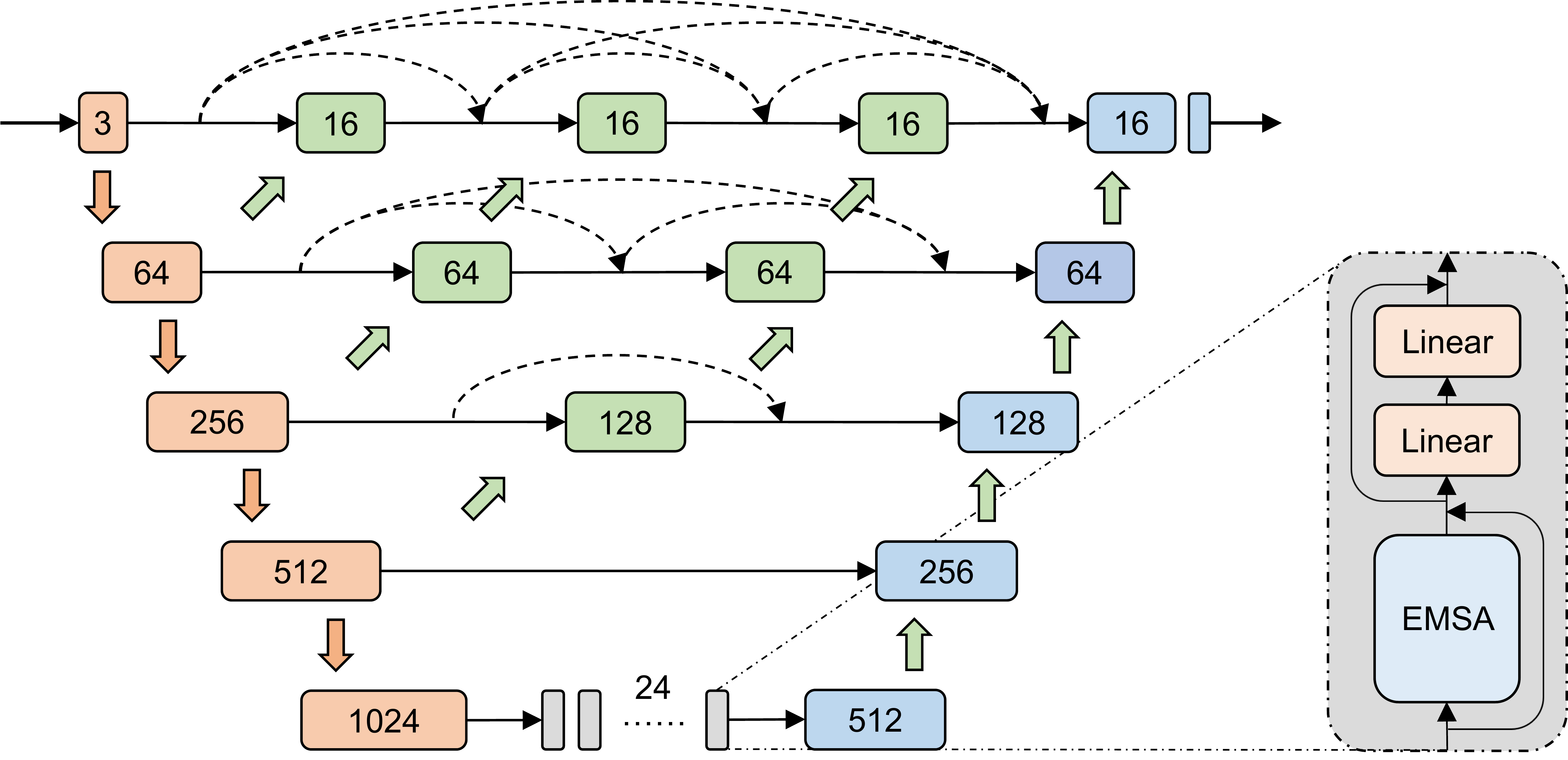}
	  \caption{The structure of the MS-TransUNet++ \citep{wang2022multiscale}. The representative TransUNet++ implements the transformer blocks into the U-Net intuitively and uses multiple skip-connection to bridge the features at different resolutions. The orange, green, and blue blocks illustrate CNN-based blocks. The gray blocks represent the transformer layers consisting of efficient multi-head self-attention (EMSA). The orange arrows show the downsampling operation, the green arrows represent the upsampling operation, and the dotted lines show the skip connection.}
	  \label{fig12}
\end{figure*}

\textbf{Multiple}. A large part of the work for coping with several modalities is U-shaped or its variants \citep{huang2022missformer, hong2022dual, lin2022contrans, huang2022scaleformer, sagar2022emsvit, jiang2022transcunet, li2022cats, wu2022d, zhang2022apt, azad2022transnorm, du2022swinpa, li2022transformer, feng2023convwin, zhang2023pct, lin2022ds, cai2022dstunet, wang2022mixed, sagar2021vitbis, huang2022tdd, park2022swine, dhamija2022semantic, pham2022segtransvae, valanarasu2021medical, tang2023combined, yuan2023effective, wang2022multiscale, wang2022net, saeed2022tmss, li2021x}. Yuan et al. \citep{yuan2023effective} developed a CTC-Net, where two encoders are designed by the swin transformer and residual CNN to capture complementary features. The cross-domain fusion block is used to concatenate these features. The correlation between features from the ResNet and transformer domains is calculated and channel attention is employed to extract dual attention information. A feature complementary module is constructed by incorporating cross-domain fusion, feature correlation, and dual attention. Zhang et al. developed an ST-Unet \citep{zhang2023st} which leverages the swin transformer to extract features. Features of each encoder stage are then enhanced by the developed CLFE module and concatenated with the current ones, followed by the up-sampling. The CLFE utilizes the self-attention block to learn the global feature information of a certain layer and fuse and learn the information with the tokens of the previous layer to obtain the enhanced multi-layer feature information. Cross-layer features are finally obtained for decoding feature enhancement. The segmented results across ST-Unet and other models can be found in \Cref{fig11}. It can be seen that for images with clear visual and semantic relationships, U-Net, TransUNet, and ST-Unet exhibit accurate segmentation. However, ST-Unet outperforms other methods for images with discreet visual relationships due to better global context encoding and semantic discrimination, and other methods can perform over- and under-segmentation. It is worth noting that due to insufficient semantic information and blurred boundaries, sometimes all the abovementioned methods cannot produce outstanding results. An example of this would be the pancreas in the first and third rows. Wang and co-workers \citep{wang2022multiscale} designed a representative architecture MS-TransUNet++. The MS-TransUNet++ implements the transformer blocks into the U-Net intuitively and uses multiple skip-connection to bridge the features at different resolutions. We show the structure of the MS-TransUNet++ in \Cref{fig12} for a more intuitive explanation. The MS-TransUNet++ has a U-shape and several constructed transformer layers are inserted between the encoder and decoder. Skip connections on different feature scales are implemented densely across different CNN blocks. The O-Net introduced in \Cref{tab1} can also segment images using a separate output head, as shown in \Cref{fig9}. The end-to-end multimodal TMSS network \citep{saeed2022tmss} proposed in 2022 is composed of a transformer encoder and CNN decoder. The transformer encoder takes the projected features from multimodal images and electronic health records.

There are a few works that are not based on U-Net \citep{azad2022transdeeplab, wang2023self, xu2021ect, jiang2022self, jiang2022transformer}. For instance, Jiang et al. \citep{jiang2022self} proposed an SMIT method to perform self-supervised learning for the transformer. The proposed method combines a dense pixel-wise regression pretext task with masked patch token distillation. Two transformers are utilized in the proposed method, serving as student and teacher, respectively. The parameters of teacher network parameters are updated using an exponential moving average with momentum. In 2022, Jiang and co-workers \citep{jiang2022transformer} introduced a progressive sampling transformer, in which a gated position-sensitive axial attention mechanism is introduced in the attention module. Iterative sampling for sampling position updating is also added to ensure the attention stays on the region to be segmented. 

\subsection{Miscellaneous}
\label{4.3}

\begin{table}
\centering
\caption{Transformer-based miscellaneous applications.}
\resizebox{\linewidth}{!}{
\begin{tabular*}{610pt}{ m{2.4cm}<{\centering} m{5.4cm}<{\centering} c m{1.8cm}<{\centering} m{3.6cm}<{\centering} m{4.8cm}<{\centering} }
\toprule
    Task                     & Method                                                         & Year  & Modality         & Object                             & Dataset \\
\midrule 
    Captioning               & transformer \citep{mohsan2022vision}                           & 2022  & X-ray            & chest                              & \citep{demner2016preparing} \\
    Captioning               & CEDT \citep{lee2022cross}                                      & 2022  & X-ray            & chest                              & chest disease \citep{wang2017chestx}, \citep{demner2016preparing} \\
    Captioning               & RATCHET \citep{hou2021ratchet}                                 & 2021  & X-ray            & chest                              & \citep{johnson2019mimic} \\
    Captioning               & TranSQ \citep{kong2022transq}                                  & 2022  & X-ray            & chest                              & \citep{demner2016preparing, johnson2019mimica} \\
    Captioning               & multicriteria supervised transformer \citep{wang2022automated} & 2022  & X-ray            & chest                              & \citep{demner2016preparing, johnson2019mimica} \\
    Captioning               & CGT \citep{li2022cross}                                        & 2022  & DFI              & retinal                            & \citep{li2021ffa} \\
    Captioning               & KdTNet \citep{cao2022kdtnet}                                   & 2022  & Endoscopy, X-ray & gastrointestinal, chest            & \citep{demner2016preparing} \\
    Captioning               & SGT \citep{lin2022sgt}                                         & 2022  & Endoscopy        & kidney, small intestine            & \citep{allan20202018} \\
    Captioning               & MCGN \citep{wang2022medical}                                   & 2022  & X-ray            & chest                              & \citep{johnson2019mimica} \\
    Captioning               & Eddie-Transformer \citep{nguyen2022eddie}                      & 2022  & X-ray            & chest                              & chest disease \citep{wang2017chestx}, \citep{demner2016preparing}, lung disease \citep{cohen2020covida} \\
    Registration             & TD-Net \citep{song2022td}                                      & 2022  & MRI              & brain                              & Alzheimer's \citep{marcus2007open} \\
    Registration             & SymTrans \citep{ma2022symmetric}                               & 2022  & MRI              & brain                              & Alzheimer's \citep{marcus2007open} \\
    Registration             & TransMorph \citep{chen2022transmorph}                          & 2022  & MRI, CT          & brain, chest, abdomen, pelvis      & \citep{iximri, segars2013population} \\
    Registration             & FTNet \citep{hu2022fusing}                                     & 2022  & MRI              & brain                              & Alzheimer's \citep{marcus2007open}, \citep{lpba40} \\
    Registration             & Swin-VoxelMorph \citep{zhu2022swin}                            & 2022  & MRI              & brain                              & Alzheimer's \citep{mueller2005ways}, Parkinson's disease \citep{marek2011parkinson} \\
    Registration             & Xmorpher \citep{shi2022xmorpher}                               & 2022  & CT               & cardiac                            & \citep{zhuang2016multi, gharleghi2022automated} \\
    Registration             & C2FViT \citep{mok2022affine}                                   & 2022  & MRI              & brain                              & Alzheimer's \citep{marcus2007open}, \citep{shattuck2008construction} \\
    Detection                & swin transformer \citep{betancourt2023transformer}             & 2023  & X-ray            & breast                             & \citep{halling2020optimam} \\
    Detection                & DETR-based \citep{leng2023deep}                                & 2023  & Microscope       & cell                               & \citep{kouzehkanan2022large} \\
    Detection                & NucDETR \citep{obeid2022nucdetr}                               & 2022  & Microscope       & cell                               & \citep{graham2019hover}, cancer \citep{gamper2020pannuke} \\
    Detection                & lightweight transformer \citep{zhang2022lightweight}           & 2022  & X-ray            & breast                             & breast cancer \citep{buda2020data} \\
    Detection                & MS Transformer \citep{shou2022object}                          & 2022  & CT               & lung, bone, kidney, lymph          & pulmonary nodules, bone lesions, kidney lesions, lymph node enlargement \citep{yan2018deeplesion} \\
    Detection                & SFOD-Trans \citep{liu2022sfod}                                 & 2022  & CT               & vein                               & \citep{everingham2015pascal} \\
    Detection                & federated split transformer \citep{park2021federated}          & 2021  & X-ray            & lung                               & lung disease \citep{covidrsna} \\
    Enhancement              & SSTrans-3D \citep{xie2022deep}                                 & 2023  & CT               & brain                              & - \\
    Enhancement              & 3D CVT-GAN \citep{zeng20223d}                                  & 2022  & PET              & brain                              & - \\
    Enhancement              & GVTrans \citep{korkmaz2021deep}                                & 2021  & MRI              & brain                              & \citep{iximri, fastmri} \\
    Enhancement              & RSTUnet-CR \citep{zhuang2022residual}                          & 2022  & US               & breast                             & - \\
    Enhancement              & TED-Net \citep{wang2021ted}                                    & 2021  & CT               & liver                              & metastatic lesion \citep{mccollough2017low} \\ 
    Enhancement              & SIST \citep{yang2022low}                                       & 2022  & CT               & head, chest, abdomen, spine, lung  & acute cognitive or motor deficit, high-risk for pulmonary nodules, metastatic liver lesions \citep{moen2021low} \\
    Enhancement              & Eformer \citep{luthra2021eformer}                              & 2022  & CT               & liver                              & metastatic lesion \citep{mccollough2017low} \\
    Localization             & transformer graph network \citep{viti2022transformer}          & 2022  & CT               & artery                             & coronary plaque \citep{zreik2018recurrent} \\
    Synthesis                & ResViT \citep{dalmaz2021resvit}                                & 2021  & MRI, CT          & brain, pelvis                      & \citep{iximri}, brain tumor \citep{menze2014multimodal}, \citep{nyholm2018mr} \\
\bottomrule
\end{tabular*}
}
\label{tab5}
\end{table}

Miscellaneous works are discussed in a sequence of captioning, registration, detection, enhancement, localization, and synthesis. It is worth noting that we do not further divide the included works by different imaging modalities due to the small number of works. The summary of the transformer-based miscellaneous applications can be found in \Cref{tab5}.

\begin{figure*}
	\centering
		\includegraphics[width=0.92\textwidth]{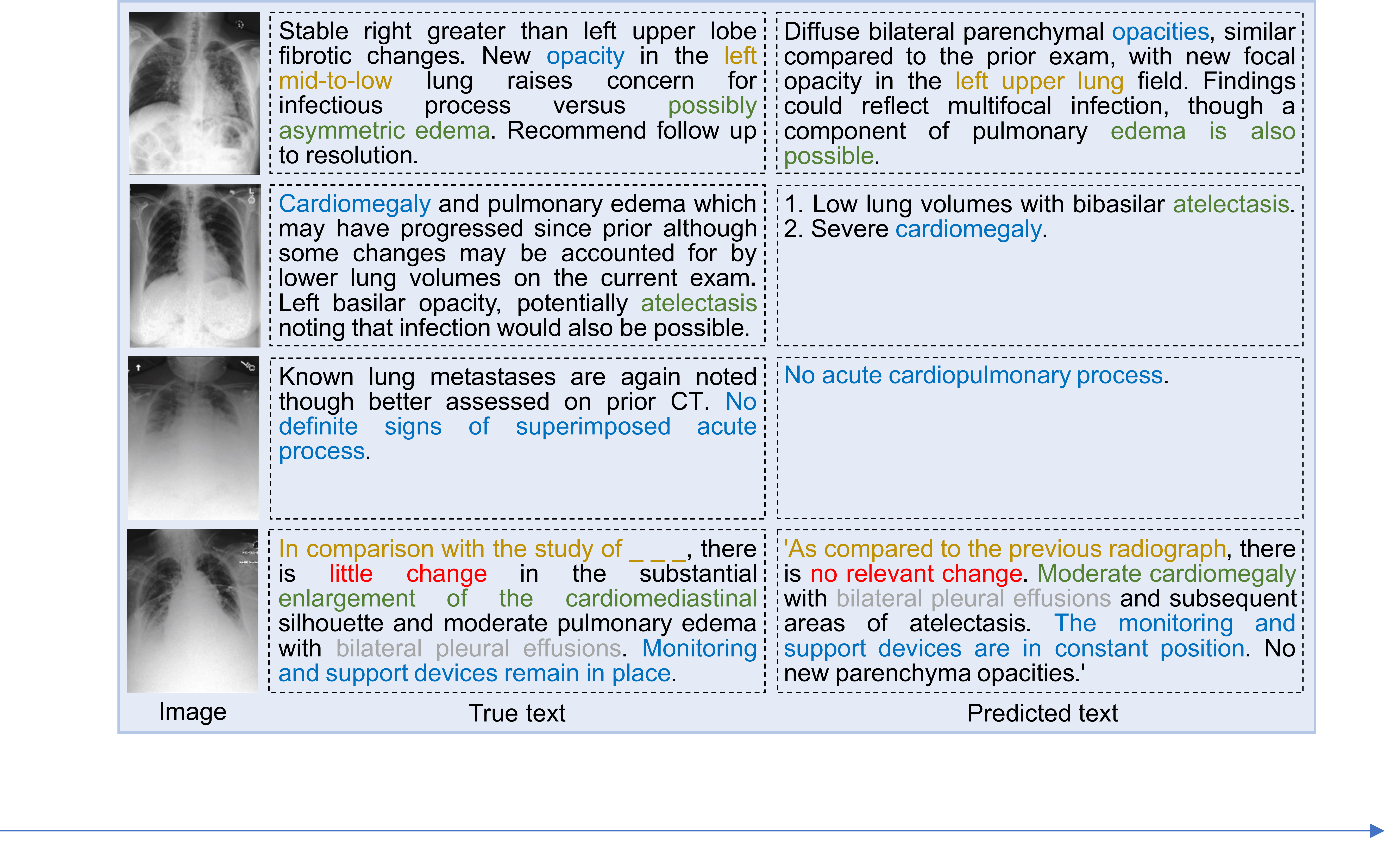}
	  \caption{Reports generated by RATCHET \citep{hou2021ratchet} using the X-ray dataset \citep{johnson2019mimic}. The same color shows the corresponding descriptions.}
	  \label{fig13}
\end{figure*}

\textbf{Captioning}. Most captioning works modify different parts of the transform from NLP \citep{mohsan2022vision, lee2022cross, hou2021ratchet, kong2022transq, wang2022automated, li2022cross, cao2022kdtnet}. For instance, Hou et al. \citep{hou2021ratchet} proposed a RATCHET network, in which the image is fed into a DenseNet-based encoder and its output passes the masked multi-head attention module. The text tokens are fed for embedding and then pass the transformer decoder. Their method performs well and is capable of generating correct keywords for the given images, as shown in ~\Cref{fig13}. Li and co-workers \citep{li2022cross} proposed a CGT network, which can restore a sub-graph from clinical relation. The restored triples are injected into the visual features as prior knowledge to drive the decoding procedure. Then, the visible matrix is utilized to limit the impact of knowledge during encoding. Reports are predicted by the encoded cross-modal features via a transformer decoder. In 2022, KdTNet \citep{cao2022kdtnet} is developed, in which the visual grid and graph convolutional modules are designed to extract fine-grained visual features. The transformer decoder is implemented to generate the hidden semantic states. A BERT-based auxiliary language module is used to obtain the context language features from the pre-defined medical term knowledge. Besides, a multimodal information fusion module is constructed to calculate the contribution of linguistic and visual features adaptively. Several works construct the models using smaller units. For example, Lin et al. \citep{lin2022sgt} designed an SGT network, in which relation-driven attention is proposed to facilitate the interaction described in the report. Instead of directly leveraging the inputs traditionally, relation-driven attention utilizes diverse sampled interactive relationships as augmented memory. Besides, an ingenious approach is also developed to homogenize the input heterogeneous scene graph, in which graph-induced attention is injected into the encoder for local interactions encoding. Wang and co-workers \citep{wang2022medical} proposed an MCGN method, in which a memory-augmented sparse attention block with bilinear pooling is developed for extracting higher-order interactions. The Eddie-Transformer developed by Nguyen et al. \citep{nguyen2022eddie} decouples the latent visual features into semantic disease embeddings and disease states using the proposed state-aware mechanism. The learned diseases and corresponding states are entangled into explicit and precise disease representations.

\textbf{Registration}. U-shaped networks are used in most of the works for medical image registration \citep{song2022td, ma2022symmetric, chen2022transmorph, hu2022fusing, zhu2022swin, shi2022xmorpher}. For example, Shi and co-workers \citep{shi2022xmorpher} developed a new transformer architecture XMorpher with dual parallel feature extraction networks. The XMorpher exchanges information through cross-attention to discover multi-level semantic correspondence. At the same time, respective features are captured gradually for registration. The cross-attention transformer blocks can find the correspondence automatically and prompt the feature fusion. There is a work that uses a different way to design the model. Mok et al. \citep{mok2022affine} proposed a C2FViT method, which naturally leverages the global connectivity and locality of the convolutional transformer and the multi-resolution strategy to learn the global affine registration. The proposed C2FViT is divided into three stages to the affine registration in a coarse-to-fine manner. The three stages have an identical architecture, including a convolutional patch embedding layer and several transformer encoder blocks. For any transformer encoder block, it is composed of an alternating multi-head self-attention module with a convolutional feed-forward layer.

\textbf{Detection}. Several works made minor modifications based on existing models for medical image detection \citep{betancourt2023transformer, leng2023deep}. One example would be the modified DETR-based \citep{carion2020end} proposed by Leng et al. \citep{leng2023deep}. The authors introduce the PVT \citep{wang2021pyramid} and deformable attention module into the DETR. Connecting existing models in series is also widely used \citep{obeid2022nucdetr, zhang2022lightweight, shou2022object}. For instance, Zhang and co-workers \citep{zhang2022lightweight} proposed a lightweight transformer for tumor detection \citep{buda2020data}, in which images are fed into a ResNet to generate feature maps. The proposed method employs attention to the outputs of ResNet to improve the hidden representations. The outputs are then fed to FPN \citep{lin2017feature}, where the multi-scale pyramidal hierarchy is utilized to construct feature pyramids. A semi-supervised method is also introduced in the detection task. Liu et al. \citep{liu2022sfod} proposed a semi-supervised framework SFOD-Trans, consisting of two parallel branches. The two branches in the SFOD-Trans are utilized to train supervised and unsupervised loss, respectively. The combination of the two branches results in a semi-supervised loss. Besides, a new fusion module named normalized ROI fusion (NRF) is designed for fusing the hepatic portal vein information captured from labeled and unlabeled images. The NRF extracts the ROI of the object region by calculating the geometric gravity center of the bounding box using real and artificial labels. The obtained two ROIs are fused using the MixUp \citep{zhang2017mixup} method. The federated split transformer discussed in \Cref{tab1} and \Cref{tab3} can also be used for image detection.

\begin{figure*}
	\centering
		\includegraphics[width=0.92\textwidth]{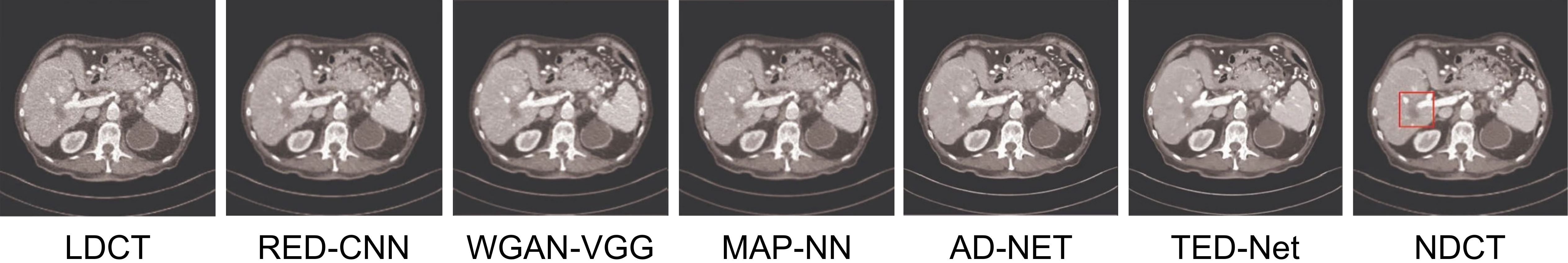}
	  \caption{The denoising results of different methods on the CT dataset \citep{mccollough2017low}. Included methods are RED-CNN \citep{chen2017low}, WGAN-VGG \citep{yang2018low}, MAP-NN \citep{shan2019competitive}, AD-NET \citep{tian2020attention}, and TED-Net. LDCT represents low-dose CT, and NDCT illustrates normal-dose CT.}
	  \label{fig14}
\end{figure*}

\textbf{Enhancement}. Reconstruction is one of the most widely researched enhancement tasks. Xie and co-workers \citep{xie2022deep} proposed a network SSTrans-3D, which reconstructs the volume using a slice-by-slice scheme. The structures of the encoder and decoder are the same as the ones in the transformer, while the normalization layers are removed. In 2022, Zeng et al. \citep{zeng20223d} developed a 3D convolutional visual transformer-GAN model 3D CVT-GAN. A hierarchical generator is designed where multiple 3D CVT blocks are used as the encoder and TCVT blocks are implemented as the decoder. The CVT block is based on convolutional embedding and the transformer, while the TCVT block is based on transpose convolutional embedding as well as the transformer. The discriminator is also based on the 3D CVT block. Korkmaz and co-workers \citep{korkmaz2021deep} proposed a novel GVTrans deep generative network. It realizes scan-specific reconstruction by embedding visual converters into the generative network. The multi-layer architecture increases image resolution progressively. Upsampled feature maps are fed into a cross-attention transformer module within each layer and generated images are masked with the same sampling pattern as in the undersampled acquisition. Besides, the parameters of the model are optimized for consistency. The RSTUnet-CR discussed in \Cref{tab2} is also capable of restructuring images through a consistency regularization decoder. Besides reconstruction, denoising is also widely investigated. Wang et al. \citep{wang2021ted} proposed a symmetric TED-Net network, consisting of an encoder-decoder structure and both the encoder and decoder contain several transformer blocks. The input to the encoder is tokenized and the decoder outputs the detokenized result. A transformer block is employed to link the encoder and decoder and the input is removed from the output to calculate the final result. The denoising results of TED-Net and other methods can be found in \Cref{fig14}, and it is easy to find that the TED-Net is capable of keeping high-level smoothness and details when removing the artifact or noise, while other methods left more blotchy noise. Yang and co-workers \citep{yang2022low} developed a SIST method, in which denoising is performed in the sinogram and image domains using the internal structure in the sinogram domain. In detail, the CT imaging mechanism and statistical characteristics of sinogram are studied for inner-structure loss design to restore high-quality CT images. A sinogram transformer module is also proposed, in which interrelations between projections of different view angles are exploited for sinogram denoising. Moreover, an image reconstruction module is developed to denoise complementarily in both the sinogram and the image domain. The Eformer \citep{luthra2021eformer} developed by Luthra et al. is built based on transformer blocks with non-overlapping window-based self-attention. The learnable Sobel-Feldman operators are incorporated to enhance edges and concatenated in the intermediate layers.

\textbf{Localization}. Viti et al. \citep{viti2022transformer} developed a transformer graph network, which exploited the self-attention mechanism of the spatial transformer to embed the contextual features of the coronary tree. Specifically, the local features are extracted by CNN and the positional encodings are then embedded into the extracted features. Positional encodings are locally calculated by utilizing the directed tree structure. A simple signed hop count from the center node is utilized, that is, +1 for distal and -1 for proximal. The resulting features are merged within the self-attention block of the spatial transformer. Besides, 150 coronary CT angiography scans are collected retrospectively.

\textbf{Synthesis}. The ResViT model proposed by Dalmaz et al. \citep{dalmaz2021resvit} is composed of an encoder, an information bottleneck, and a decoder. The generator of the ResViT utilizes a central bottleneck with aggregated residual transformer (ART) blocks with transformer modules. The ART block is composed of three parts, which are the transformer encoder-based part, the channel compression part, as well as the residual CNN part. For the given input feature maps, it first passes the transformer encoder-based part, in which the residual connection is implemented. The concatenated features are then fed to the channel compression part with two CNN branches followed by the sum operation. Output feature maps are obtained by feeding the output of the channel compression part to the residual CNN part. As the name implies, the residual connection is also implemented in this block.


\subsection{Quantitative Evaluation}
\label{4.4}

\Cref{tab6} shows the performance of the representative transformer-based models and the performance comparison with other state-of-the-art methods. For classification \citep{krishnan2021vision, tamhane2022colonoscopy, li2022crossa, tummala2022breast, gul2022histopathological, al2022covid}, optimal accuracy, F1 score, area under the curve, precision, recall, balanced accuracy, and Matthew’s correlation coefficient are observed. For accuracy, we observe an accuracy of up to 99.6\% with the leadership of up to 5.6\%. Regarding the F1 score, the highest F1 score of 99.5\% and the highest leadership of 3.1\% are reached. As for area under the curve, precision, recall, balanced accuracy, and Matthew’s correlation coefficient, the highest results are 99.4\%, 99.5\%, 98.8\%, 99.4\%, and 98.9\%, respectively. For segmentation \citep{junayed2022transformer, alahmadi2022semi, gao2021utnet, luo2021ucatr, liu2022ccat, ning2021cac}, superior dice similarity coefficient, sensitivity, specificity, pixel accuracy, and intersection over union are achieved. The highest dice similarity coefficient is 90.6\% and the highest improvement reaches 4.4\%. As for sensitivity, the highest values and leadership are 94.8\% and 9.6\%, respectively. Regarding the specificity, pixel accuracy, and intersection over union, the best results are 97.7\%, 95.9\%, and 80.7\% respectively. For captioning \citep{cao2022kdtnet}, superior bilingual evaluation understudy-4, consensus-based image description evaluation, and recall-oriented understudy for gisting evaluation-longest common subsequence are reached, equaling 0.58, 0.69, and 0.75, respectively. As for registration \citep{song2022td} and enhancement \citep{wang2021ted, yang2022low}, better dice similarity coefficient, structural similarity index, root mean squared error, and peak signal-to-noise ratio of 74.3\%, 0.92, 8.77, and 41.80 are observed.

As can be seen, the reviewed transformer-based method outperforms most existing methods on different MIA tasks. These methods include both the CNN-based methods \citep{li2021dual, silva2020covid, al2019deepa, kirillov2017unified, yin2019domain, huang2017densely, he2016deep} and the transformer-based methods \citep{woo2018cbam, liu2021exploring, mok2020fast, tian2020attention, chen2021transunet, liu2021swin}. Overperforming CNN-based models can prove the inherent advantage of the transformer-based method while outperforming existing transform-based methods illustrates the effectiveness of their proposed improvements. The outstanding results prove the versatility and adaptability of the transformer-based method in the field of MIA. Though the transformer-based method mostly outperforms existing methods to a large extent, we need to point out that there exist some cases that it is not well-at. For example, the transformer-based method can sometimes get a relatively low specificity. To sum up, with overall satisfactory performance, the development of MIA can be significantly boosted by the transformer-based method.

\begin{table}
\centering
\caption{Quantitative performance of the representative transformer-based method. For classification, ACC, F1, AUC, PRE, REC, BA, and MCC represent the accuracy, F1 score, area under the curve, precision, recall, balanced accuracy, and Matthew’s correlation coefficient, respectively. For segmentation, DSC, SEN, SPE, PA, and IoU stand for dice similarity coefficient, sensitivity, specificity, pixel accuracy, and intersection over union, respectively. For miscellaneous tasks, BLEU-4, CIDEr, ROUGE-L, DSC, SSIM, and RMSE, PSNR mean bilingual evaluation understudy-4, consensus-based image description evaluation, recall-oriented understudy for gisting evaluation-longest common subsequence, dice similarity coefficient, structural similarity index, root mean squared error, and peak signal-to-noise ratio, respectively. We select one of the representative datasets when multiple datasets are used.}
\resizebox{\linewidth}{!}{
\begin{tabular*}{604pt}{ m{2.4cm}<{\centering} m{5.2cm}<{\centering} m{3cm}<{\centering} m{9cm}<{\centering}}
\toprule
    Task           & Method                                                      & Baseline                                  & Performance (Baseline) \\
\midrule 
    Classification & transformer \citep{krishnan2021vision}                      & DenseNet                                  & ACC: 97.6\% (92.0\%), F1: 94.6\% (91.5\%), PRE: 95.3\% (91.0\%), REC: 93.8\% (92.2\%) \\
    Classification & transformer \citep{tamhane2022colonoscopy}                  & ResNet                                    & ACC: 81.8\% (73.1\%) \\ 
    Classification & multi-scale feature fusion transformer \citep{li2022crossa} & DenseNet                                  & ACC: 85.3\% (84.3\%), F1: 74.2\% (72.7\%), AUC: 92.3\% (90.7\%), PRE: 80.2\% (80.3\%), REC: 70.5\% (68.7\%) \\
    Classification & ensembled swin transformer \citep{tummala2022breast}        & swin transformer                          & ACC: 99.6\% (99.2\%), F1: 99.5\% (99.2\%), AUC: 99.4\% (99.2\%), BA: 99.4\% (99.1\%), MCC: 98.9\% (98.3\%) \\
    Classification & Self-ViT-MIL \citep{gul2022histopathological}               & DSMIL \citep{li2021dual}                  & ACC: 91.5\% (91.5\%), AUC: 94.3\% (93.6\%) \\
    Classification & symmetric dual transformer \citep{al2022covid}              & EfficientNet-based \citep{silva2020covid} & ACC: 99.1\% (99.0\%), F1: 99.1\% (99.0\%), PRE: 99.5\% (99.2\%), REC: 98.8\% (98.8\%) \\
    Segmentation   & versatile transformer \citep{junayed2022transformer}        & FrCN \citep{al2019deepa}                  & DSC: 85.3\% (82.0\%), SEN: 83.9\% (80.8\%), SPE: 82.4\% (83.5\%), PA: 88.1\% (85.3\%), IoU: 80.7\% (77.4\%) \\
    Segmentation   & semi-supervised transformer \citep{alahmadi2022semi}        & MSA-UNet \citep{alahmadi2022multiscale}   & DSC: 90.6\% (90.3\%), SEN: 94.8\% (88.7\%), SPE: 97.7\% (97.1\%), PA: 95.9\% (95.8\%) \\
    Segmentation   & UTNet \citep{gao2021utnet}                                  & CBAM \citep{woo2018cbam}                  & DSC: 88.3\% (87.3\%) \\
    Segmentation   & UCATR \citep{luo2021ucatr}                                  & TransUNet                                 & DSC: 73.6\% (70.6\%), SEN: 73.1\% (69.4\%) \\
    Segmentation   & CCAT-net \citep{liu2022ccat}                                & FPN \citep{kirillov2017unified}           & DSC: 65.1\% (60.7\%), SEN: 76.0\% (66.4\%), SPE: 97.7\% (95.5\%) \\
    Segmentation   & CAC-EMVT \citep{ning2021cac}                                & TransUNet                                 & DSC: 75.4\% (73.2\%), PA: 94.0\% (92.3\%), IoU: 80.6\% (78.0\%) \\
    Captioning     & KdTNet \citep{cao2022kdtnet}                                & PPKED \citep{liu2021exploring}            & BLEU-4: 0.58 (0.58), CIDEr: 0.69 (0.68), ROUGE-L: 0.75 (0.74) \\
    Registration   & TD-Net \citep{song2022td}                                   & SYMNet \citep{mok2020fast}                & DSC: 74.3\% (73.7\%) \\
    Enhancement    & TED-Net \citep{wang2021ted}                                 & AD-Net                                    & SSIM: 0.91 (0.90), RMSE: 8.77 (9.72) \\
    Enhancement    & SIST \citep{yang2022low}                                    & DP-ResNet \citep{yin2019domain}           & SSIM: 0.92 (0.91), PSNR: 41.80 (40.92) \\
\bottomrule
\end{tabular*}
}
\label{tab6}
\end{table}


\section{Challenges and Perspectives}
\label{5}

Despite the significant progress and successful deployment of transformer-based methods as a major game changer in the CV area of MIA, future challenges still exist. We summarize several main challenges and give corresponding perspectives on how to solve or improve them. We organize the main challenges together with the corresponding perspectives into three parts, which are feature integration and computing cost reduction, data augmentation and dataset collection, and learning manner and modality-object distribution.

\textbf{Feature integration and computing cost reduction}. In order to improve the model performance by capturing both local and global features, most current works only simply hybridize CNN and transformer, such as inserting the transformer encoder block into a CNN model. However, the integration of local and global features in this way may not be firm enough. To integrate CNN and the transformer closer, two-fold ways can be implemented by benefiting the transformer from inductive biases, which are inherent in CNN. On the one hand, inductive bias in CNN can be brought back to the transformer \citep{zhang2023vitaev2, chen2021chasing, yuan2021tokens}. On the other hand, the transformer can learn with CNN simultaneously under the mutual learning framework \citep{zhang2022bootstrapping}. High computing cost is always an inevitable problem for the transformer due to the quadratic computational complexity of the input size, especially when the image resolution is high. However, seldom works mentioned or try to solve this problem. To improve the training efficiency of the transformer, more attention computing methods, such as shifted window attention \citep{liu2021swin}, efficient attention \citep{shen2021efficient}, and multi-head linear self-attention \citep{wang2020linformer} can be taken into consideration. Besides, projection parameters in the transformer can be shared at different levels. The FLOPs and the number of parameters of the model can be calculated for quantitative model complexity evaluation and further comparison.

\textbf{Data augmentation and dataset collection}. In the field of MIA, data shortage always hamper the model performance. The data augmentation technique is an important research direction to address this problem. However, as far as we have seen, many of the transformer-related works have not gone deep into it. Most of the works only use traditional data augmentation techniques such as rotation, crop, and flip. So far, only seldom works utilize advanced data augmentation methods, such as the GAN-based method to synthesize images. Even though, the implemented basic GAN cannot be considered advanced as the quality and resolution of the images synthesized by basic GAN are difficult to guarantee. In the case of using low-quality or even repetitive (e.g., model collapse) synthesized images for training, the validity of the model performance is questionable. For instance, a classification model can show very high accuracy on a dataset, but there may exist thousands of repetitive synthesized images that are correctly classified in the dataset. To augment data better, state-of-the-art image synthesis models should be taken into consideration. For instance, GAN that suits small datasets like StyleGAN2-ADA \citep{karras2020training}, independent spatial and appearance transform models \citep{zhao2019data}, and diffusion probabilistic models like 3D-DDPM \citep{dorjsembe2022three}. Another problem we observed is that many selected papers only compare the model performance with several classic models, and models designed for MIA by other authors are not included. This is especially common for non-mainstream modalities and objects. One of the main reasons that caused it is the lack of widely accepted benching marking datasets like ImageNet \citep{deng2009imagenet}. Thus, the collection and publication of new high-quality medical datasets can benefit this research field a lot. Constructing such datasets can also benefit the development of the transfer learning technique in the MIA field. According to our observation, though transfer learning is widely implemented in the field of MIA, most of them transfer from ImageNet. As natural images and medical images can have different data contributions, transferring from medical datasets may further improve model performance.

\textbf{Learning manner and modality-object distribution}. There are several state-of-the-art learning manners, such as weakly-supervised learning, and unsupervised learning, which can reduce the need for data labeling. However, these manners are not widely used in transformer-based MIA works. Regarding the modality-object distribution, most existing works mainly concentrate on several mainstream modalities, as shown in \Cref{fig8}. However, there is a lot of untapped research potential outside of these mainstream modalities and objects. In terms of modalities, current research primarily focuses on MRI, CT, X-ray, and microscope imaging. Despite being an essential medical image modality, the US has not been fully investigated. In terms of objects, most of the current works focus on the brain, chest, abdomen, and heart, while other objects such as the retina warrant further investigation.

We believe that future efforts in the transformer-based MIA community are certainly not limited to the three points listed above. More research directions like model interpretability should also be fully investigated. With the joining of more and more artificial intelligence and medical researchers, the transformer-based MIA will be developed at an unprecedented speed from both algorithm and data sides. Besides self-benefiting, the fast development of the transformer-based MIA can also benefit related application domains of MIA a lot. For example, state-of-the-art methods \citep{su2022multilevel, qi2022directional, hu2022colorectal} in these related application domains such as the optimization algorithm can be combined with the transformer. Specifically, the optimization algorithm can be implemented to search the hyperparameter combination in the transformer model to search for further performance improvement. With the fast development of the transformer-based method, the development of MIA can definitely be accelerated to a large extent. This can help doctors to diagnose more fastly, accurately, and smartly, promoting early intervention.


\section{Conclusion}
\label{6}

The transformer-based MIA is now developing rapidly. In this review, we summarize and analyze the recent progress on transformer-based MIA. The structure of this review is on the basis of different tasks, including classification, segmentation, captioning, registration, detection, enhancement, localization, and synthesis. The task-modality mode can make the readers access their needs faster and easier. We also compare the performance between the transformer-based method and existing state-of-the-art methods. Furthermore, we point out the current challenges and perspectives in the transformer-based MIA field in three core points from both data and algorithm sides. The main advantages of our task-modality review include updated content, detailed information, and comprehensive comparison. Our systematic review may help new DL researchers as well as medical experts without DL knowledge enter this field more quickly. In other words, the detailed content about the latest progress and the performance comparison can be easily accessed with the task-modality mode. However, it is worth noting that the task-modality organization mode may occasionally overlook the sequential relationship between research works. Specifically, subsequent studies building on prior works may be categorized into different tasks or modalities due to the tasks performed or datasets used. While this is not likely to happen frequently, it can occur in certain cases. We show that future work of the transformer-based MIA can be five-fold. First, the method exploration for feature integration and computing cost reduction can be developed. Second, more effort on data augmentation as well as dataset collection should be paid. Next, more focus on the non-mainstream learning manner, modality, and object is needed. Then, deeper research on the model interpretability can be performed. Finally, the transformer-based method and other related application domain methods can be combined. To sum up, benefiting from the fast development of the transformer-based MIA, the medical diagnosis might become more and more convenient and accurate.



\setlength{\bibsep}{0.0pt}
\bibliographystyle{unsrt}
\bibliography{reference.bib}
\biboptions{sort&compress}







\end{sloppypar}
\end{document}